\documentclass[pra,superscriptaddress,nofootinbib,showkeys, preprint]{revtex4}
\usepackage[usenames,dvipsnames]{color}
\usepackage{mathtools}
\usepackage[colorlinks]{hyperref}
\usepackage{float}
\usepackage{graphicx}
\usepackage[latin1]{inputenc}
\usepackage{amsmath}
\usepackage{amsfonts}
\usepackage{amssymb}
\usepackage{graphicx}
\DeclareGraphicsExtensions{.pdf,.png,.jpg}

\usepackage{epstopdf}
\def\beq{\begin{equation}}
\def\eeq{\end{equation}}
\def\bea{\begin{eqnarray}}
\def\eea{\end{eqnarray}}

\begin{document}
		\title{  Super Periodic Potential}
			\author{ Mohammd Hasan}
	\email{mhasan@isac.gov.in, \ \ mohammadhasan786@gmail.com} 
	\affiliation{ISRO Satellite Centre (ISAC),
Bangalore-560017, INDIA }
		
		\author{Bhabani Prasad Mandal}
			\email{bhabani.mandal@gmail.com}
			\affiliation{Department of Physics, Institute of Science, Banaras Hindu University, 
				 Varanasi - 221005, India}
		\begin{abstract}

In this paper we introduce the concept of super periodic potential (SPP) of arbitrary order $n$, $n \in I^{+}$, in one dimension. General theory of wave propagation through SPP of order $n$ is presented and the reflection and transmission coefficients are derived in their closed analytical form by transfer matrix formulation. We present scattering features of super periodic rectangular potential and super periodic delta potential as special cases of SPP. It is found that the symmetric self-similarity is the special case of super periodicity. Thus by identifying a symmetric fractal potential as special cases of SPP, one can obtain the tunnelling amplitude for a particle from such fractal potential.  By using the formalism of SPP we obtain the close form expression of tunnelling amplitude of a particle for general Cantor and Smith-Volterra-Cantor potentials.     

\end{abstract}

\maketitle

\section{Introduction}

There is always a recurring interest, both theoretical and experimental, in the study of one dimensional scattering by finite periodic chains of non-overlapping barriers or wells. This problem has its own significance because it exhibits important features of quantum mechanics: tunnelling and interference \cite{hwlee, tmk} which serves practical motivations for the physics of lattices and superlattices in solid state physics and electronic devices \cite{fcap}, optical  multiquantum well systems \cite{neidlinger}, and theory of X-ray reflection from amorphous superlattices \cite{rhan}. The historic paper of Kronig and Penney (1930) , describes motion of an electron in an infinite (or half-infinite) one-dimensional periodic chain of delta- potentials \cite{kroning}, and has served as the most important tool in explaining various interesting physical properties of real materials. Similar problem with a finite number of scattering centres has also been studied. Pshenichnov (1962) has investigated the scattering problem on a periodic finite chain of identical potentials using the WKB approximation and has shown the existence of  resonance effect in the transmission coefficient \cite{psh}. Reading and Siege(1972) have considered particle scattering from a finite chain of delta- potentials of arbitrary strengths and positions using momentum representation method \cite{reading}. The scattering problem for a finite chain of $N$ equally spaced symmetric potentials has been studied by many groups in twentieth century by various methods and formalisms \cite{mcv,kalotas}. Scattering from finite and infinite repeated structures to see the band-structure characteristics of periodic potentials along with resonances are also observed \cite{hwlee,griffith, rozman, sprung}.   

However, the realization of scattering and resonances from a super periodic potential with multiple periodicities which is the most appropriate and generalized representation of superlattices are still phenomenological and there is no concrete theoretical analysis present in literature, to the best of our knowledge. Superlattices are in general a periodic structure of layers of two or more materials, typically having the thickness of each layer of several nano-metres.  Many artificially created superlattices have been proposed with various solid state systems with utmost two or three patterns of repetitions and some are also experimentally studied \cite{15}. It will be extremely helpful to have the generalised version of scattering matrices and coefficients for any multiple patterns of repetition to make more and more advanced superlattices. In parallel, some super-periodic structures are also invented in optics to serve practical purposes in modern science and technologies \cite{leong2011,leong2012} with two or three periodicities, but without having any analytical generalised form of the scattering matrices. No attempt has been made yet for constructing the generalised form of scattering matrices for a super-periodic potential with arbitrary number periodicities. Of course this problem is quite challenging because of its extremely complicated structure and analytical lengthiness. In this work we not only develop the most generalised structure of super-periodic potentials but also explicitly construct the corresponding transfer matrices and scattering coefficients in the closed forms. 

Our aim in this article is to present the exact scattering solutions and resonances for such an array of quantum mechanical potentials which is distributed in a finite or infinite space with a super-periodic manner. This  super periodic potential of order `n'  is described as follows: consider a `unit cell potential' of width $2a$ periodically repeats $N_{1}$ times with the constant distance between two successive cells as $c_{1}$, which is its periodic distance. This periodic system `as a whole' repeats $N_{2}$ times with periodic distance $c_{2}$. This new system again wholly repeats $N_{3}$ times with periodic distance $c_{3}$.  This process of periodic repetition, of a new potential system each time, continues upto an arbitrary number of times $N_{n}$ with the periodic distance $c_{n}$, where n is positive integer. The closed form expression of the scattering coefficients (reflection and transmission) for such a generalised super periodic distribution of several sets of potentials is constructed in this work. We treat the problem of order $n$ in terms of the transfer matrix for  each unit cell potential. Specific examples of super periodic potential are discussed when unit cell potential is delta and rectangular potential.

It is an extraordinary fact to see that a family of symmetric fractal potentials is the special cases of super periodic potential. This shows that the concept of super periodicity is more general than the periodicity and being the super set of a family of symmetric potential broaden its scope to great length. Thus identifying a fractal potential as super periodic potential completes the theory of tunnelling amplitude from such potential in analytical form. There have been extensive works on the tunnelling problem of potentials which has fractal distributions \cite{f1}-\cite{f8}. Much of these works have relied on the numerical multiplication of the transfer matrices as well as their analytical properties. We identify the one dimensional symmetric Cantor potential of stage $G$, namely the standard and general Cantor, Smith-Volterra-Cantor potential as special cases of super periodic rectangular potential of order $G$. The close form expressions of these fractal potentials of arbitrary stage $G$ are provided based on the developed formulation of SPP.   

We organize the paper as follows. In section \ref{general_theory} we briefly discuss the general theory of wave propagation in a locally periodic media. This is a brief representation of the work presented in \cite{griffith_pp}. In section \ref{general_theory_spp} we present our generalization of locally periodic potential to locally super periodic potential. Here we explicitly derive the expressions of scattering coefficients of wave scattered from one dimensional SPP. The expression of transmission amplitude is presented in the nice closed form. The general features of scattering from SPP are discussed in section \ref{general_features}. The special cases of SPP for delta and rectangular potential as `unit cell potential' are discussed in section \ref{special_cases}. In this section we also present the symmetric Cantor potentials as special cases of SPP. Finally we discuss our results in section \ref{results_discusssions}.   

\section{General Theory: One dimensional scattering of Schrodinger Particle}

\label{general_theory}
\begin{figure}
\begin{center}
\includegraphics[scale=0.5]{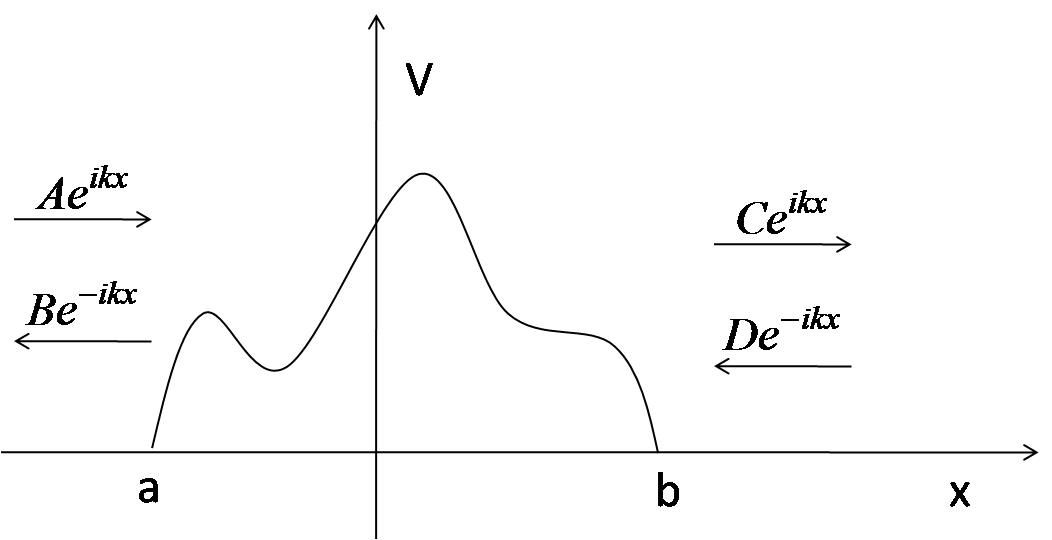}
\caption{Wave scattering from an arbitrary potential}
\label{spotential}
\end{center}
\end{figure}

In this section we discuss the transfer matrix method of one dimensional scattering of Schrodinger particle from a  potential barrier. For a potential  $V(x)$, $a<x<b$ as shown in Fig \ref{spotential}, the time independent Schrodinger equation for a particle of mass `$m$' and energy `$E$' is
\begin{equation}
\label{tise}
-\frac{ \hbar^{2}}{2m}\frac{d^2}{dx^2}\psi(x) +V(x)\psi(x)=E \psi(x)
\end{equation}  
The general solution of Eq. (\ref{tise}) in all three regions are 
\begin{eqnarray}
\psi(x)&=& A e^{ikx}+ B e^{-ikx}  \ \ \ \ \ \mbox{for} \ x<a  \nonumber  \\
\psi (x)&=&\psi_{ab}(x) \ \ \ \ \ \ \ \ \ \  \ \ \ \ \ \ \mbox {if } a<x<b  \nonumber \\
\psi(x) &=& Ce^{ikx}+ D e^{-ikx} \ \ \ \ \ \mbox{for} \  x>b.
\end{eqnarray} 
where $k=\frac{\sqrt{2mE}}{\hbar}$. These solutions along with the time factor $e^{\frac{iEt}{\hbar}}$ represent the waves travelling to left and right of the barrier. The coefficients $A$, $B$, $C$ and $D$ are the amplitude of the waves. This is shown graphically in Fig. \ref{spotential}.   The solution of Eq. \ref{tise} along with the boundary conditions of the wave function $\psi(x)$ , the continuity of $\psi(x)$ and its first derivative at the boundary `a' and `b', give two linear equations involving the coefficients $A$, $B$, $C$ and $D$. This equation in matrix form would be
\[ \begin{pmatrix}   A \\ B   \end{pmatrix} =M \begin{pmatrix}   C \\ D   \end{pmatrix}\] 
Here `$M$' is a $2 \times 2$ matrix.
\begin{equation}
 M= \begin{pmatrix}   m_{11} & m_{12} \\ m_{21} & m_{22}   \end{pmatrix}  
\label{transfer_matrix_unit_cell_elements}
\end{equation}
 \\
Due to the time reversal invariance of the Schrodinger equation along with the conservation of probability, it has been shown that for all Hermitian potential $V(x)$, the elements of transfer matrix $M$ have following form \cite{griffith_pp},
\begin{subequations}
\begin{equation}
m_{11}=m_{22}^{*}
\end{equation}
\begin{equation}
m_{12}=m_{21}^{*}
\end{equation}
\label{transfer_matrix_unit_cell}
\end{subequations}
and $m_{11}$ and $m_{12}$ satisfy
\begin{equation}
|m_{11}|^{2}-|m_{12}|^{2}=1
\end{equation}
Thus the transfer matrix $M$ is unimodular. In terms of the elements of the transfer matrix $M$ for the potential $V(x)$ the reflection and transmission coefficients are
\begin{equation}
t_{l}=t_{r}=\frac{1}{m_{22}}
\end{equation} 
\begin{equation}
r_{l}=\frac{m_{21}}{m_{22}} , 
r_{r}=\frac{m_{12}}{m_{22}}
\end{equation}

\subsection{Generalization to locally periodic potential}

\begin{figure}
\begin{center}
\includegraphics[scale=0.5]{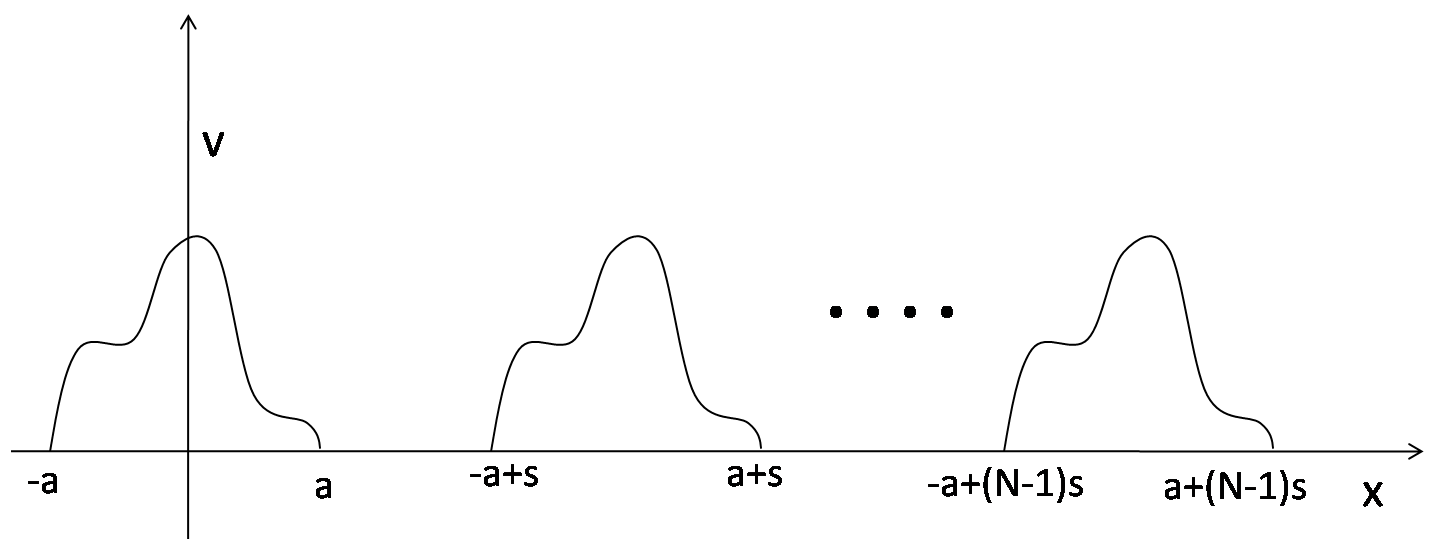}
\caption{A locally periodic potential}
\label{periodicpotential}
\end{center}
\end{figure}
We will extensively use the formalism developed by Griffiths et. al. \cite{griffith_pp} for the scattering of waves from a locally periodic media. Therefore we briefly discuss the theory here. Consider the periodic potential shown in Fig. \ref{periodicpotential}. The potential $V(x)$, called as `unit cell' potential , periodically repeats `$N$' times over the distance `$s$', $s \geq 2a$ (see Figure \ref{periodicpotential}). If the transfer matrix of the unit cell potential is known (Eq. \ref{transfer_matrix_unit_cell_elements}), then the transfer matrix for the  whole periodic system ( array of N unit cell potential ) is 
\begin{equation}
 M_{N}=\begin{pmatrix} [m_{11} e^{-iks}U_{N-1}(\xi)-U_{N-2}(\xi)]e^{ikNs} & m_{12}U_{N-1}(\xi)e^{-ik(N-1)s} \\  m_{12}^{*}U_{N-1}(\xi)e^{ik(N-1)s}  &  [m_{11}^{*} e^{iks}U_{N-1}(\xi)-U_{N-2}(\xi)]e^{-ikNs}   \end{pmatrix} 
\label{transfer_matrix_periodic_potential}
\end{equation} 
where $U_{N}(\xi)$ is the Chebyshev polynomial of second kind and the argument $\xi$ is 
\begin{eqnarray}
\xi &=& \frac{1}{2} (m_{11}e^{-iks}+m_{22}e^{iks}) \nonumber \\
	&= & \mbox{Re}(m_{22})\cos{ks}-\mbox{Im}(m_{22})\sin{ks} .
\label{xi_expression1}
\end{eqnarray}
Since $m_{22}=|m_{22}|e^{i\alpha}=m_{11}^{*}$ where $\alpha$ is the argument of $m_{22}$ $\xi$  can be written as
\begin{equation}
\xi = |m_{22}|\cos{(\alpha+ks)}
\label{xi_expression2}
\end{equation}
In terms of the elements of the matrix $M$( Eq. \ref{transfer_matrix_unit_cell_elements}), the elements of $M_{N}$ are
\begin{subequations}
\begin{equation}
(M_{N})_{11} =[m_{11} e^{-iks}U_{N-1}(\xi)-U_{N-2}(\xi)]e^{ikNs} 
\label{m11_griffith}
\end{equation}
\begin{equation}
(M_{N})_{12}= m_{12}U_{N-1}(\xi)e^{-ik(N-1)s}
\label{m12_griffith}
\end{equation}
\begin{equation}
(M_{N})_{21}= m_{21}U_{N-1}(\xi)e^{ik(N-1)s}
\label{m21_griffith}
\end{equation}
\begin{equation}
(M_{N})_{22}=[m_{22} e^{iks}U_{N-1}(\xi)-U_{N-2}(\xi)]e^{-ikNs} 
\label{m22_griffith}
\end{equation}
\end{subequations} 
Now the transmission coefficient for the periodic system is
\begin{equation}
t_{l}=t_{r}=\frac{1}{(M_{N})_{22}}
\end{equation} 
By using the unimodular property of the transfer matrix $M$, the transmission probability, $T=\vert t_{l} \vert ^{2}=\vert t_{r} \vert ^{2}$, can be written as 

\begin{equation}
T=|t_{l}|^2=|t_{r}|^2=\frac{1}{1+(|m_{11}|U_{N-1}(\xi))^{2}}
\label{transmission_N}
\end{equation} \\
Expressing the Chebyshev polynomials of the second kind in terms of sinusoidal functions,
\begin{equation}
U_{N}(\xi)=\frac{\sin(N+1)\gamma}{\sin{\gamma}}
\end{equation}  
the transmission probability is obtained in trigonometric form as
\begin{equation}
T=\left[ 1+ |m_{11}|^{2}\Big (\frac{\sin{N\gamma}}{\sin{\gamma}}\Big )^{2}\right]^{-1}
\end{equation}  

where 
\begin{equation}
\gamma=\cos^{-1}\xi \ .
\end{equation}

\section{Waves in Super Periodic Potential}
\label{general_theory_spp}
\begin{figure}
\begin{center}
\includegraphics[scale=0.5]{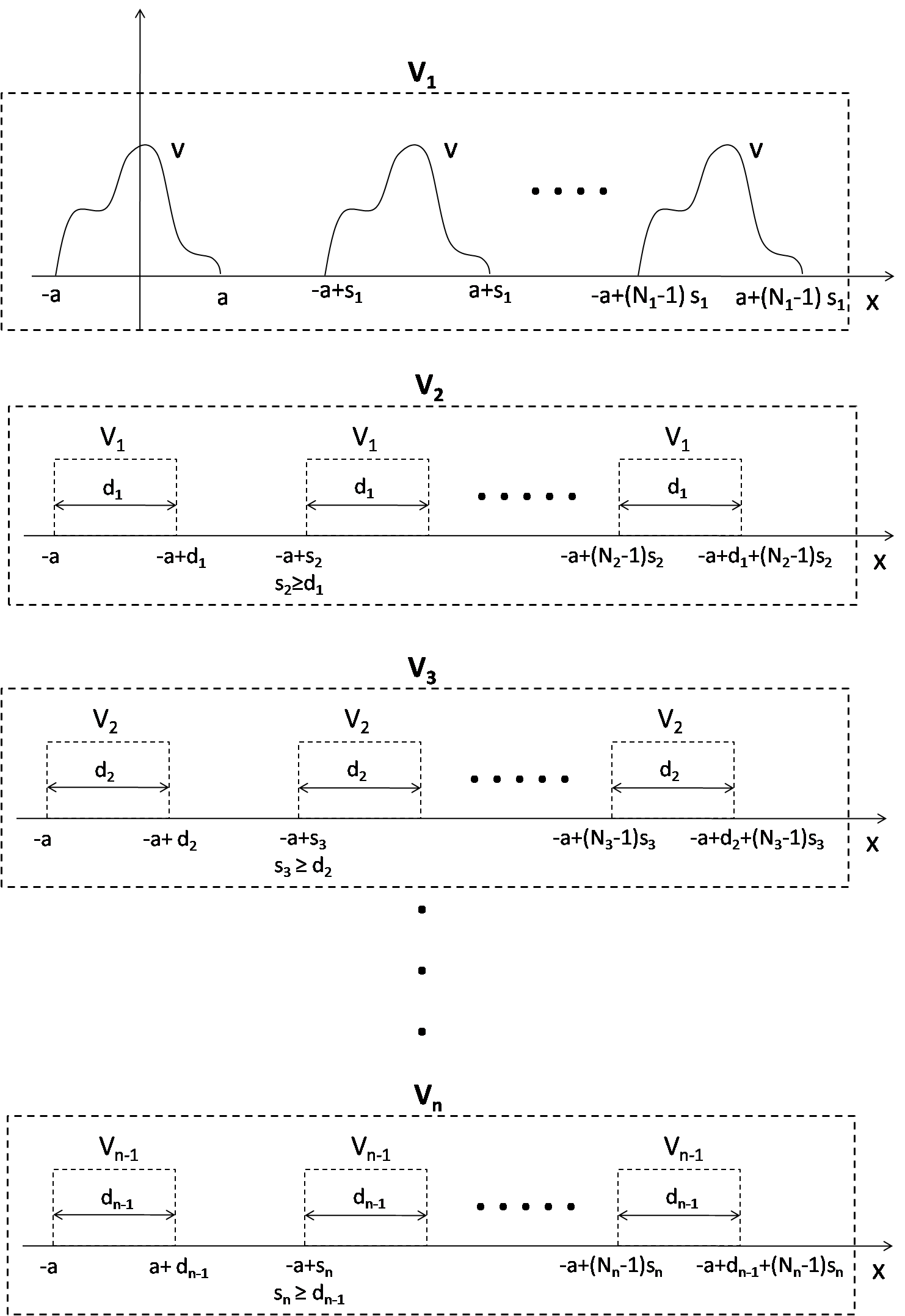}
\end{center}
\caption{Illustration of a super periodic potential}
\label{superperiodicpotential}
\end{figure}
The super periodic potential is illustrated in  Fig. \ref{superperiodicpotential}. Consider a unit cell potential $V(x),\ -a<x<a$, of span $2a$ which repeats periodically $N_{1}$ times. The cells are separated with distance $s_{1}\geq 2a$ as shown in top of  Fig. \ref{superperiodicpotential}. The total span of this periodic potential is $d_{1}=2a+(N_{1}-1)s_{1}$. Let us call this whole set as a new potential represented as $V_{1}$. The potential cell $V_{1}$ is replicated $N_{2}$ times at equal intervals $s_{2} \geq d_{1}$. We again call this new set as $V_{2}$ (of span $d_{2}=d_{1}+(N_{2}-1)s_{2}$) and replicate it again at equal intervals $s_{3} \geq d_{2}$.  This process of replicating new system every time at regular intervals continues till an arbitrary number $n$ such that we obtain super periodic potential $V_{n}$ of total span $d_{n}=d_{n-1}+(N_{n}-1)s_{n}$, $n \in I^{+}$, as shown in Fig \ref{superperiodicpotential}. This is super periodic potential of order $n$ of the potential $V(x),\ -a<x<a$. We wish to obtain the close form expression of transfer matrix of potential $V_{n}$ in terms of the transfer matrix of unit cell potential $V(x)$ (Eq. (\ref{transfer_matrix_unit_cell_elements}) or Eq. (\ref{transfer_matrix_unit_cell})). From the transfer matrix we can further obtain the close form expressions of scattering amplitudes for the super periodic potential $V_{n}$. The calculation is detailed below.
\subsection{Transfer Matrix of potential $V_{n}$}
We denote the elements of the transfer matrix of super periodic potential $V_{n}$ as $(m_{11})_{n}$, $(m_{12})_{n}$, $(m_{21})_{n}$ and $(m_{22})_{n}$. The calculation of diagonal and off-diagonal elements of the transfer matrix is presented below.
\subsubsection{Calculation of $(m_{11})_{n}$ and $(m_{22})_{n}$}
We  first calculate the element $(m_{22})_{n}$. The potential $V=V(x)$ acts as a `unit cell' for potential $V_{1}$. Therefore $(m_{22})_{1}$ is obtained by using Eq. \ref{m22_griffith} as
\begin{equation}
(m_{22})_{1}=m_{22} e^{-ik(N_{1}-1)s_{1}}U_{N_{1}-1}(\xi_{1})-U_{N_{1}-2}(\xi_{1})e^{-ikN_{1}s_{1}}
\label{m22_1}
\end{equation}
The arguments of Chebyshev Polynomial (of second kind) $\xi_{1}$ for $V_{1}$, $\xi_{2}$ for $V_{2}$ and similarly $\xi_{n}$ for $V_{n}$ will be evaluated later.
Now $V_{1}$ act as a `unit cell' for potential $V_{2}$. Therefore the $m_{22}$ element for potential $V_{2}$ is
\begin{equation}
(m_{22})_{2}=(m_{22})_{1} e^{-ik(N_{2}-1)s_{2}}U_{N_{2}-1}(\xi_{2})-U_{N_{2}-2}(\xi_{2})e^{-ikN_{2}s_{2}}
\label{m22_2}
\end{equation}
Similarly $V_{2}$ act as `unit cell' for potential $V_{3}$ and hence,
\begin{equation}
(m_{22})_{3}=(m_{22})_{2} e^{-ik(N_{3}-1)s_{3}}U_{N_{3}-1}(\xi_{3})-U_{N_{3}-2}(\xi_{3})e^{-ikN_{3}s_{3}}
\label{m22_3}
\end{equation}
With the same logic one arrives at,
\begin{equation}
(m_{22})_{n-1}=(m_{22})_{n-2} e^{-ik(N_{n-1}-1)s_{n-1}}U_{N_{n-1}-1}(\xi_{n-1})-U_{N_{n-1}-2}(\xi_{n-1})e^{-ikN_{n-1}s_{n-1}}
\label{m22_n_1}
\end{equation}
and,
\begin{equation}
(m_{22})_{n}=(m_{22})_{n-1} e^{-ik(N_{n}-1)s_{n}}U_{N_{n}-1}(\xi_{n})-U_{N_{n}-2}(\xi_{n})e^{-ikN_{n}s_{n}}
\label{m22_n}
\end{equation}
Sequentially using values of $(m_{22})_{n-1}$, $(m_{22})_{n-2}$, ......,$(m_{22})_{3}$, $(m_{22})_{2}$, $(m_{22})_{1}$ in Eq. \ref{m22_n} and after performing laborious algebra, we obtain the compact form expression of $(m_{22})_{n}$ as,
\begin{equation}
(m_{22})_{n}=m_{22} e^{-ik \sum_{p=1}^{n} (N_{p}-1)s_{p}} \prod_{p=1}^{n} U_{N_{p}-1}(\xi_{p})-\sum_{r=1}^{n-1}L_{r} -U_{N_{n}-2}(\xi_{n})e^{-ikN_{n}s_{n}}
\label{m22_nf}
\end{equation}
where,
\begin{equation}
L_{r}=e^{-ik ( \sum_{p=1}^{n} N_{p}s_{p} -\sum_{p=1}^{r} N_{p-1}s_{p-1})}e^{ik ( \sum_{p=1}^{n} s_{p} -\sum_{p=1}^{r} s_{p})} U_{N_{r}-2}(\xi_{r}) \prod_{p=r+1}^{n} U_{N_{p}-1}(\xi_{p})
\label{lr}
\end{equation}
with $N_{0}=0$, $s_{0}=0$. \\
By the same logic used in $(m_{22})_{n}$, one can easily obtain the expression for $(m_{11})_{n}$. This is given below
\begin{equation}
(m_{11})_{n}=m_{11} e^{ik \sum_{p=1}^{n} (N_{p}-1)s_{p}} \prod_{p=1}^{n} U_{N_{p}-1}(\xi_{p})-\sum_{r=1}^{n-1}G_{r} -U_{N_{n}-2}(\xi_{n})e^{ikN_{n}s_{n}}
\label{m11_nf}
\end{equation}
where,
\begin{equation}
G_{r}=e^{ik ( \sum_{p=1}^{n} N_{p}s_{p} -\sum_{p=1}^{r} N_{p-1}s_{p-1})}e^{-ik ( \sum_{p=1}^{n} s_{p} -\sum_{p=1}^{r} s_{p})} U_{N_{r}-2}(\xi_{r}) \prod_{p=r+1}^{n} U_{N_{p}-1}(\xi_{p})
\label{gr}
\end{equation}
$N_{0}$ and $s_{0}$ have been defined earlier. It is easy to see that $L_{r}=G_{r}^{*}$.  Combining this with $m_{11}=m_{22}^{*}$, we find $(m_{11})_{n}=(m_{22})_{n}^{*}$.
\subsubsection{Calculation of $(m_{12})_{n}$ and $(m_{21})_{n}$}
Since potential  $V$ is `unit cell' potential for $V_{1}$, therefore we use Eq. \ref{m12_griffith} to calculate $(m_{12})_{1}$. This is similar process as noted earlier.
\begin{equation}
(m_{12})_{1}=m_{12}U_{N_{1}-1}(\xi_{1})e^{-ik(N_{1}-1)s_{1}}
\label{m12_1}
\end{equation}
Similarly,
\begin{equation}
(m_{12})_{2}=(m_{12})_{1}U_{N_{2}-1}(\xi_{2})e^{-ik(N_{2}-1)s_{2}}
\label{m12_2}
\end{equation}
and so on,
\begin{equation}
(m_{12})_{n-1}=(m_{12})_{n-2}U_{N_{n-1}-1}(\xi_{n-1})e^{-ik(N_{n-1}-1)s_{n-1}}
\label{m12_n_1}
\end{equation}
and,
\begin{equation}
(m_{12})_{n}=(m_{12})_{n-1}U_{N_{n}-1}(\xi_{n})e^{-ik(N_{n}-1)s_{n}}
\label{m12_n}
\end{equation}
Sequentially substituting the values of $(m_{12})_{n-1}$, $(m_{12})_{n-2}$, . . ., $(m_{12})_{2}$, $(m_{12})_{1}$ in Eq. \ref{m12_n}, we get the expression for $(m_{12})_{n}$.
\begin{equation}
(m_{12})_{n}=m_{12} e^{-ik \sum_{p=1}^{n} (N_{p}-1)s_{p}} \prod_{p=1} ^{n} U_{n_{p}-1} (\xi _{p})
\label{m12_nf}
\end{equation}  
Similar calculation for $(m_{21})_{n}$ results,
\begin{equation}
(m_{21})_{n}=m_{21} e^{ik \sum_{p=1}^{n} (N_{p}-1)s_{p}} \prod_{p=1} ^{n} U_{n_{p}-1} (\xi _{p})
\label{m21_nf}
\end{equation}  
As $m_{12}=m_{21}^{*}$, we see from Eq. \ref{m12_nf} and \ref{m21_nf} that $(m_{12})_{n}=(m_{21})_{n}^{*}$. 
\subsubsection{Calculation of $\xi_{n}$}
From Eq. \ref{xi_expression1} we understand that for potential $V_{1}$,
\begin{equation}
\xi_{1}=\mbox{Re}[m_{22}]\cos{ks_{1}}-\mbox{Im}[m_{22}]\sin{ks_{1}}  \nonumber
\end{equation}
Similarly for potential $V_{2}$, we calculate $\xi_{2}$ as follows,
\begin{equation}
\xi_{2}=\mbox{Re}[(m_{22})_{1}]\cos{ks_{2}}-\mbox{Im}[(m_{22})_{1}]\sin{ks_{2}} 
\label{xi2_1}
\end{equation}
On using Eq. \ref{m22_1} in \ref{xi2_1}, the expression for $\xi_{2}$ is simplified to
\begin{equation}
\xi_{2}=|m_{22}| U_{N_{1}-1}(\xi_{1})\cos{[\alpha-k \{(N_{1}-1)s_{1}-s_{2}\}]}-U_{N_{1}-2}(\xi_{1})\cos{\{k(N_{1}s_{1}-s_{2})\}}
\label{xi2}
\end{equation}
Performing similar algebra, we calculate
\begin{multline}
\xi_{3}=|m_{22}| U_{N_{1}-1}(\xi_{1}) U_{N_{2}-1}(\xi_{2}) \cos{[\alpha -k \{ (N_{1}-1)s_{1}+(N_{2}-1)s_{2}-s_{3}\}]}\\
-U_{N_{1}-2}(\xi_{1}) U_{N_{2}-1}(\xi_{2}) \cos{\{k(N_{1}s_{1}+N_{2}s_{2}-s_{2}-s_{3})\}}-U_{N_{2}-2}(\xi_{2})\cos{\{k(N_{2}s_{2}-s_{3})\}}
\label{xi3}
\end{multline}
We follow the same procedure to calculate $\xi_{4}$ ,$\xi_{5}$ ...$\xi_{n}$. After a lengthy algebra the general expression for $\xi_{n}$ can be obtained in the following form. 
\begin{multline}
\xi_{n}=|m_{22}|\cos{\big[ \alpha-k \{\sum_{p=1}^{n-1} (N_{p}-1)s_{p}-s_{n}\}\big]} \prod_{p=1}^{n-1}U_{N_{p}-1}(\xi_{p})-\sum_{r=1}^{n-1}H_{r} \\ -U_{N_{n-1}-2}(\xi_{n-1})\cos{k(N_{n-1}s_{n-1}-s_{n})}
\label{xin}
\end{multline}
Where,
\begin{equation}
H_{r}=\cos{\big[k\big(\sum_{p=r}^{n-1}N_{p}s_{p}-\sum_{p=r+1}^{n}s_{p}\big)\big]}U_{N_{r}-2}(\xi_{r}) \prod _{p=r+1}^{n-1}U_{N_{p}-1}(\xi_{p})
\label{hr}
\end{equation}
Therefore the transfer matrix of $\mbox{n}^{th}$ order super periodic potential $V_{n}$
\begin{equation}
 M_{n}= \begin{pmatrix}   (m_{11})_{n} & (m_{12})_{n} \\ (m_{21})_{n} & (m_{22})_{n}   \end{pmatrix}  
\label{transfer_matrix_spp}
\end{equation}
has been computed.
\subsection{Reflection and Transmission probability}
The transmission and reflection coefficient of the $\mbox{n}^{th}$ order super peridic potential is obtained from the elements of transfer matrix. 
\begin{equation}
(t_{l})_{n}=(t_{r})_{n}=\frac{1}{(m_{22})_{n}}
\label{tln}
\end{equation} 
\begin{equation}
(r_{l})_{n}=\frac{(m_{21})_{n}}{(m_{22})_{n}} , 
(r_{r})_{n}=\frac{(m_{12})_{n}}{(m_{22})_{n}}
\label{rln}
\end{equation}
In particular, the transmission probability is
\begin{equation}
T(N_{1},N_{2},...N_{n})=\frac{1}{|(m_{22})_{n}|^{2}}=\frac{1}{|(m_{11})_{n}|^{2}}
\label{transmission_1}
\end{equation}
From Eq. \ref{m12_nf} and the unimodular properties of matrix \ref{transfer_matrix_spp}  the above expression is simplified to
\begin{equation}
T(N_{1},N_{2},...N_{n})=\frac{1}{1+[|m_{12}|U_{N_{1}-1}(\xi_{1})U_{N_{2}-1}(\xi_{2})U_{N_{3}-1}(\xi_{3})........U_{N_{n}-1}(\xi_{n})]^{2}}
\label{transmission_2}
\end{equation}
Eq. \ref{transmission_2} can also be written in trigonometric form,
\begin{equation}
T(N_{1},N_{2},...N_{n};E)=\frac{1}{1+\Big(|m_{12}| \frac{\sin{N_{1}\gamma_{1}}}{\sin{\gamma_{1}}} \frac{\sin{N_{2}\gamma_{2}}}{\sin{\gamma_{2}}}............\frac{\sin{N_{n}\gamma_{n}}}{\sin{\gamma_{n}}}\Big)^{2}}
\label{transmission_3}
\end{equation}
Where,
\begin{equation} 
\gamma_{n}=\cos^{-1}{\xi_{n}}
\label{gamman}
\end{equation}
\section{General features of scattering from super periodic potential}
\label{general_features}
In this section we discuss some general features of scattering from super periodic potential. 
\subsection{Transmission Resonances} Let $E=E^{*}$ be the zeros of $m_{12}(E)$  i.e. $E^{*}$ is the  transmission resonance energy for the `unit cell' potential. Then from Eq. \ref{transmission_3} this also implies transmission resonance for super periodic potential of any order `$n$' for $E=E^{*}$. This can also be understood as follows. Since the whole super periodic system is made of `unit cell' potential as its building block, the wave at transmission resonance passes through all unit cells with perfect transmission. Hence the whole system also shows transmission resonance as well. The same logic also applies to other periodic entity that acts as unit cell for the subsequent periodic systems. If transmission resonance occurs for potential $V_{1}$ at energy $E=E_{1}^{*}$, it will also occur for the whole super periodic system at energy $E=E_{1}^{*}$. Mathematically this can be seen from Eq. \ref{transmission_3}. If for an energy, $E=E_{m_{1}}$, $N_{1}\gamma_{1}= m_{1}\pi$, $m_{1} \in \{I,0 \}$, then $T(N_{1},N_{2},...N_{n};E_{m_{1}})=1$ for $n \geq 1$, $n \in I^{+}$. Similarly if for $E=E_{m_{2}}$, $N_{2}\gamma_{2}=m_{2}\pi$, $m_{2} \in \{I,0\}$, then $T(N_{1},N_{2},...N_{n};E_{m_{2}})=1$ for $n \geq 2$, $n \in I^{+}$. The same general results apply at all those energies at which $N_{2}\gamma_{2}=m_{2}\pi$, $N_{3}\gamma_{3}=m_{3}\pi$, ........, $N_{n}\gamma_{n}=m_{n}\pi$, where $m_{2}$,$m_{3}$,.....,$m_{n}$ $\in \{I,0\}$.  
\subsection{Emergence of band structure}
The Chebyshev polynomial $U_{N}(\xi)$ diverges in the limit $N\rightarrow \infty$ when its argument $\xi$ lies outside the interval $[-1,1]$. Various $\xi_{i}$, $i \in I^{+}$ appearing as arguments of Chebyshev polynomial in the expression of transmission probability (Eq. \ref{transmission_2}) are oscillatory in nature. The presence of trigonometric terms as well as  terms of Chebyshev polynomials with arguments $\xi_{1}$, $\xi_{2}$,.....$\xi_{i-1}$ in the expression of $\xi_{i}$ causes each $\xi_{i}$ to show oscillatory behaviour with energy. Therefore there exist ranges of energy for which $\xi_{i}$ lies outside the interval $[-1,1]$. For such energy ranges the super periodic potential will be opaque for $N_{i}\rightarrow\infty$. In other words such energy ranges are forbidden for the particle inside the super periodic potential. Therefore the allowed energy of the particle will show band structure similar to that of Kroning-Penny model. However in this case the band structure will be modulated by super periodicity. This is discussed later by considering specific examples of unit cell potential.    
\section{Special cases of super periodic potential}
\label{special_cases}
In this section we discuss the special cases of unit cell potential generating the super periodic potential. The results are presented and discussed for the  delta and rectangular barrier potentials as unit cells. 
\subsection{Super periodic delta potential}
The transfer matrix for a delta potential $V(x)=V_{0}\delta(x)$ is \cite{griffith_pp}
\begin{equation}
 M= \begin{pmatrix}   1+i\beta & i\beta \\ -i\beta & 1-i\beta   \end{pmatrix}  
\label{transfer_matrix_delta_potential}
\end{equation}
where
\begin{equation}
 \beta=\frac{mV_{0}}{\hslash^{2} k}  
\label{transfer_matrix_delta_potential}
\end{equation}
\begin{equation}
k=\sqrt{2mE}
\end{equation}
Therefore for delta potential
\begin{equation}
|m_{22}|=\sqrt{1+\beta^{2}}  
\label{m22_delta}
\end{equation}
and
\begin{equation}
\alpha=-\tan^{-1}{\beta} 
\label{m22_alpha}
\end{equation}
Scattering from periodic delta potential have been discussed in details in the literatures \cite{griffith, griffith_pp}. However for the sake of completeness we provide the expression for transmission probability. The transmission probability from finite periodic delta potential (also called as Dirac comb) is
\begin{equation}
T(N_{1};k)=\frac{1}{1+[\beta U_{N_{1}-1}(\xi_{1})]^{2}} 
\label{t_delta_potential}
\end{equation}
where
\begin{equation}
\xi_{1}=\cos{ks_{1}}+\beta \sin{ks_{1}}
\end{equation}
When Dirac comb as a whole periodically repeats (locally) we have periodic Dirac comb. The transmission from periodic Dirac comb is discussed in detail below.
\subsubsection{Periodic Dirac comb}
\label{periodic_dirac_comb}
The transmission probability for the periodic Dirac comb can be straight forwardly calculated by using the Eqs. \ref{transmission_3} and \ref{xi2}.
\begin{equation}
T(N_{1},N_{2};k)=\frac{1}{1+[\beta U_{N_{1}-1}(\xi_{1})U_{N_{2}-1}(\xi_{2})]^{2}}
\label{t_periodic_Dirac_comb}
\end{equation}
where,
\begin{equation}
\xi_{2}=\sqrt{(1+\beta^{2})} U_{N_{1}-1}(\xi_{1})\cos{[\tan^{-1}{\beta}+k \{(N_{1}-1)s_{1}+s_{2}\}]}-U_{N_{1}-2}(\xi_{1})\cos{[k(N_{1}s_{1}-s_{2})]}
\label{xi2_periodic_dirac_comb}
\end{equation}
The above expression can also be written as
\begin{equation}
\xi_{2}=\sqrt{(1+\beta^{2})} U_{N_{1}-1}(\xi_{1})\cos{[\tan^{-1}{\beta}+k \{(2N_{1}-1)s_{1}+y\}]}-U_{N_{1}-2}(\xi_{1})\cos{ky}
\label{xi2_periodic_dirac_comb_other_expression}
\end{equation}
where $y=c_{2}-s_{1}$. 

\begin{figure}
\begin{center}
\includegraphics[scale=0.4]{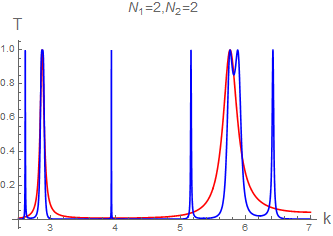} a \ \includegraphics[scale=0.4]{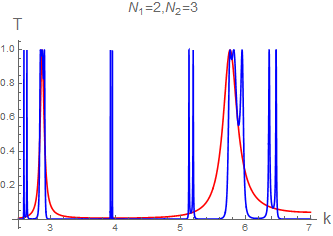} b \ \includegraphics[scale=0.4]{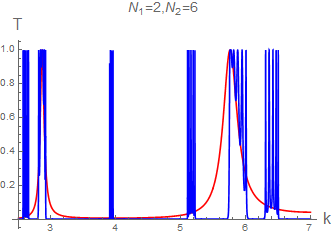} c \\  

\includegraphics[scale=0.4]{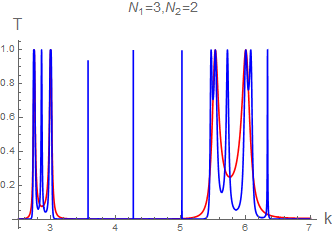} d \ \includegraphics[scale=0.4]{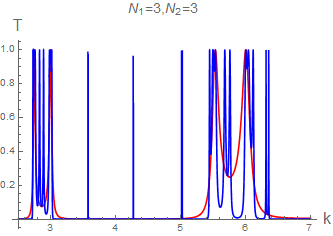} e \ \includegraphics[scale=0.4]{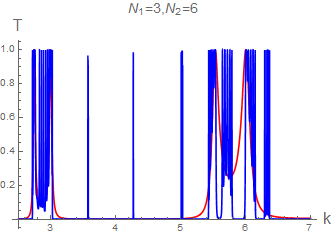} f \\
\caption{Plot showing the variation of tunnelling probability with $k$ for super periodic delta potential of order $2$. The red and blue curves are for $T(N_{1};k)$ and $T(N_{1},N_{2};k)$ respectively.  Here $V_{0}=10$, $s_{1}=1$ and $c_{2}=0.5$ and $N_{1}$, $N_{2}$ are shown in the figures.  }
\label{T_N1_N2}
\end{center}
\end{figure}
The behaviour of transmission probability for different values of $N_{1}$, $N_{2}$ are plotted in Figure \ref{T_N1_N2} with $k$. We see that the band structure appear for $N_{1}=2$, $N_{2}=2$. For all the calculations and plots in the paper  we have taken $\hbar=1$ and $m=1$. The emergence of band structure for a Dirac comb with $N_{1}=5$ has been noted earlier \cite{griffith}. In space fractional quantum mechanics, the band emerge even with $N_{1}=4$ \cite{jd_tare}.  We present a detail discussions of the transmission probability and its dependence on various parameters of Dirac comb below.
\paragraph{Resonance band:}
Consider first the case of transmission resonances. Chebyshev Polynomial $U_{N}(\xi)$ of degree $N$ has $N$ roots in the interval $[-1,1]$. Hence from Eq. \ref{t_delta_potential} we see that the Dirac comb with $N_{1}$ number of delta potentials has $N_{1}-1$ resonances in transmission for $\xi_{1}\in [-1,1]$. Let $\xi_{1}^{*}=\xi_{1}(k=k^{*})$ be one of the roots of $U_{N_{1}-1}(\xi_{1})=0$. From Eq. \ref{xi2_periodic_dirac_comb} value of $\xi_{2}^{*}=\xi_{2}(k=k^{*})$ is 
\begin{equation}
\xi_{2}^{*}=-U_{N_{1}-2}(\xi_{1}^{*})\cos{k^{*}y}
\label{xi2_star}
\end{equation} \\
 
\begin{figure}
\begin{center}
\includegraphics[scale=0.5]{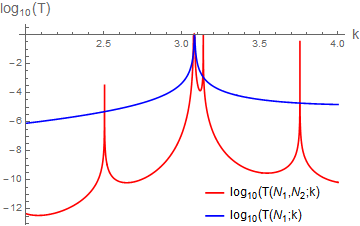}
\end{center}
\caption{Plot showing the behaviour of $T(2,2;k)$ (red curve) near the transmission resonance point $k=k^{*}=3.080068$ of   $T(2;k)$ (blue curve).  Here  $V_{0}=50$, $s_{1}=1$ and $y=c_{2}-s_{1}=4$. It is seen that the transmission resonance of $T(2;k)$ is also the transmission resonance of $T(2,2;k)$.}
\label{t1_t12_delta1}
\end{figure}

\begin{figure}
\begin{center}
\includegraphics[scale=0.4]{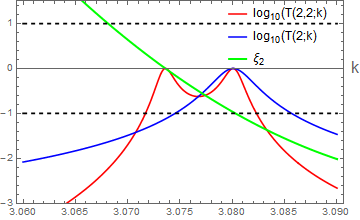} a \ \includegraphics[scale=0.4]{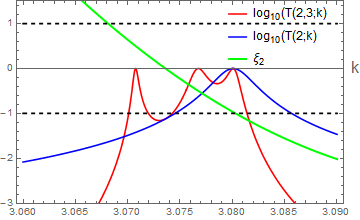} b \\
\includegraphics[scale=0.4]{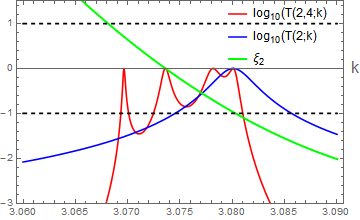} c \  \includegraphics[scale=0.4]{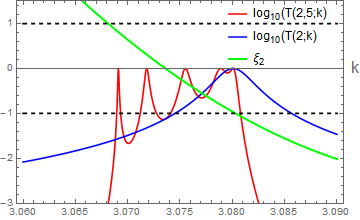} d \\
\caption{Variation of $T(2,2;k)$, $T(2,3;k)$, $T(2,4;k)$, $T(2,5;k)$ along with $\xi_{2}$ near the transmission resonance point  $k^{*}=3.080068$ of $T(2;k)$. Dotted lines are drawn at $-1$ and $+1$ to show that resonance band lies between the region where $\xi_{2} \in [-1,1]$ }
\label{periodic_Dirac_comb_near_resonance}
\end{center}
\end{figure}
\begin{figure}
\begin{center}
\includegraphics[scale=0.5]{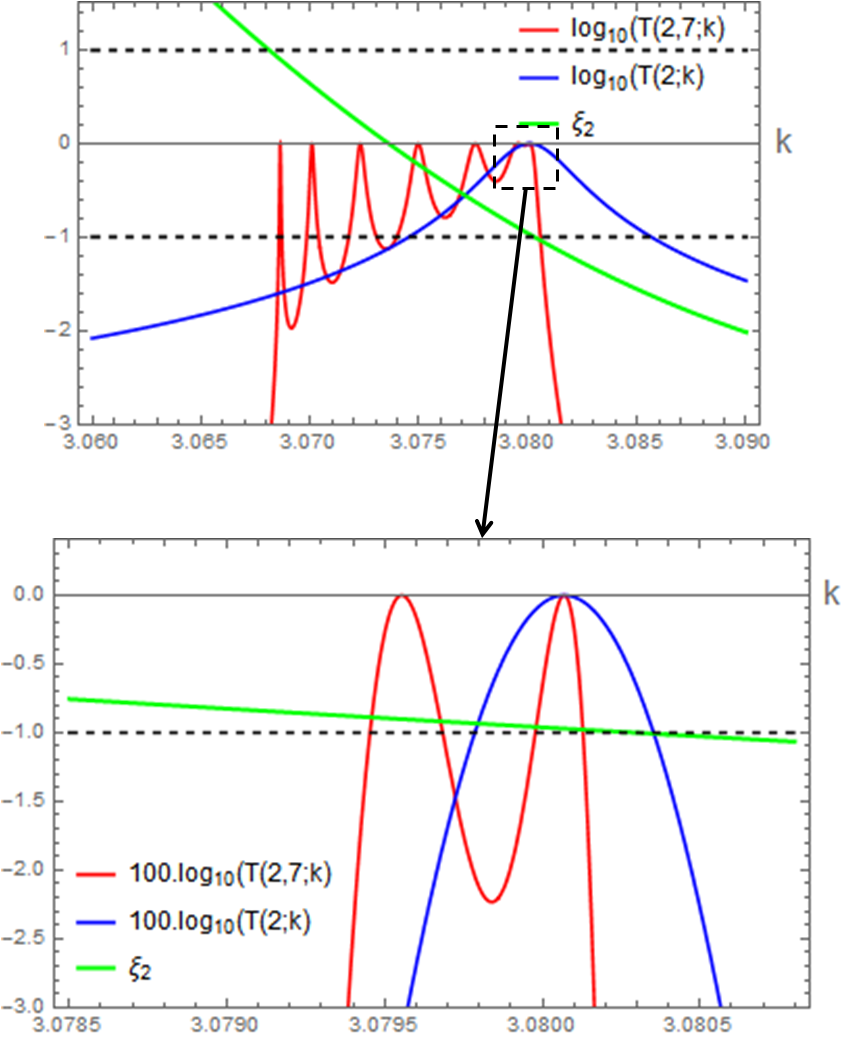}
\caption{$T(2,7;k)$ near resonance point $k^{*}=3.080068$ of $T(2;k)$ shows $7$ resonance peaks near $k^{*}$. Note that two resonance peak of $T(2,7;k)$ in the extreme neighbourhood of $k^{*}$ are difficult to resolve}
\label{T22_T27_resonance_peaks}
\end{center}
\end{figure}

\begin{figure}
\begin{center}
\includegraphics[scale=0.6]{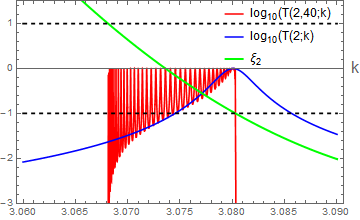}
\caption{$T(2,40 ;k)$ near resonance point $k^{*}=3.080068$ of $T(2;k)$. Other parameters of periodic Dirac comb are same as reported for Figure \ref{t1_t12_delta1}. Here the resonance band lies over the interval $\xi_{2} \in [-1,1]$}.
\label{T22_T240_resonance_peaks}
\end{center}
\end{figure}
Due to the presence of $\cos{ky}$ term in the expression of $\xi_{2}^{*}$ the value of $\xi_{2}$ in the immediate neighbourhood of $k=k^{*}$ will vary from $-1$ to $+1$. As $U_{N_{2}-1}(\xi)$ has $N_{2}-1$ roots for $\xi \in [-1,1]$ therefore in case of periodic Dirac comb, those energy points for which Dirac comb had transmission resonances will split into a `resonance band' containing $N_{2}-1$ resonance peaks. It must be noted that these $N_{2}-1$ resonance peaks will not contain the peak at $k=k^{*}$ that arises due to transmission resonance of the Dirac comb. The overall effect is that the periodic Dirac comb will show a resonance band containing the total  $(N_{2}-1)+1=N_{2}$ numbers of resonance peak in the neighbourhood of $k=k^{*}$ (and containing $k^{*}$). We found that as $N_{2}$ increases these resonance peaks become difficult to resolve and therefore increasing $N_{2}$ causes the resonance band to become more flat. We illustrate this graphically. In the Figure \ref{t1_t12_delta1} the variation of transmission probability for $T(2;k)$ and $T(2,2;k)$ is shown with $k$. Perfect transmission occur for $k=3.080068=k_{1}^{*}$. At this value of $k$, $T(2,2)$ also shows resonance features. We plot more closely near $k_{1}^{*}$ in Fig. \ref{periodic_Dirac_comb_near_resonance} and Fig.  \ref{T22_T27_resonance_peaks} for different $N_{2}$ to show the appearance of resonance band with $N_{2}$ number of resonance peaks. From these figures it is also evident that in the neighbourhood of transmission  resonance of corresponding Dirac comb, the bloch phase $\xi_{2}$ lies between $-1$ to $1$. From Figure \ref{T22_T27_resonance_peaks} it is evident that the resonance peaks become difficult to resolve as $N_{2}$ increases.
\paragraph{Width of resonance band:}    
Since these resonance peaks can only lie in the region spanned by $k$ over which $\xi_{2}\in [-1,1]$ near $k^{*}$, an increase in $N_{2}$ will cause the resonance band to effectively occupy the range of $k$ over which $\xi_{2}$ varies from $-1$ to $1$ (or vice versa). This is shown graphically in Fig. \ref{T22_T240_resonance_peaks} for $N_{2}=40$. In Fig. \ref{T22_T240_resonance_peaks} there are $40$ peaks in total in the resonance band. The width of the resonance band (measured in $k$) is the range of $k$ near $k^{*}$ over which $\xi_{2}$ varies from $-1$ to $1$. The analytical reason for this has already been discussed. Below we compute approximately the width of the resonance band. \\

\begin{figure}
\begin{center}
\includegraphics[scale=0.4]{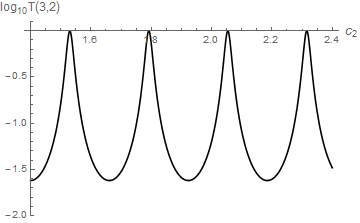} a \ \includegraphics[scale=0.4]{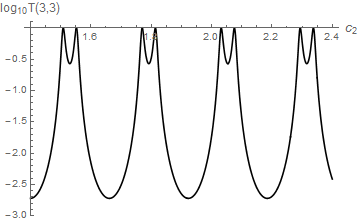} b \\
\includegraphics[scale=0.4]{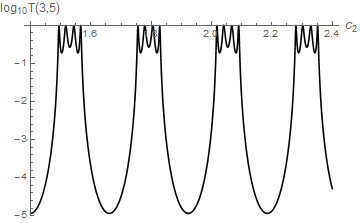} c \  \includegraphics[scale=0.4]{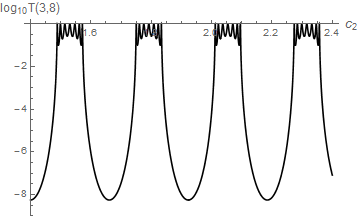} d \\
\includegraphics[scale=0.4]{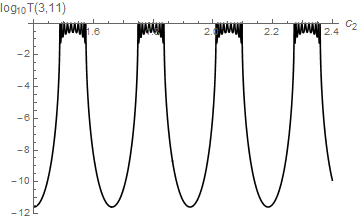} e \  \includegraphics[scale=0.4]{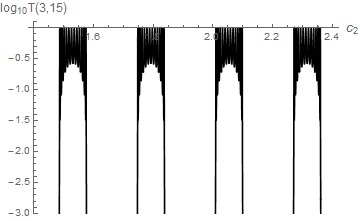} f \\
\caption{Transmission probability $T(N_{1},N_{2})$ as a function of $y$. Here $N_{1}=3$ and $N_{2}=2,3,5,8,11,15$ in $a,b,c,d,e$ and $f$ respectively. $V_{0}=20$, $s_{1}=1.0$ and $k=12$. Note the appearance of resonance band with $N_{2}-1$ numbers of resonance peaks.}
\label{TN1N2_with_y}
\end{center}
\end{figure}

\begin{figure}
\begin{center}
\includegraphics[scale=0.4]{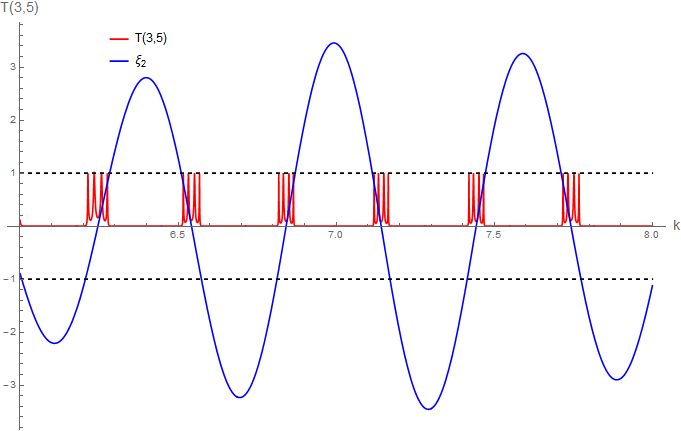} 
\caption{Transmission probability $T(3,5;k)$ and $\xi_{2}$ as a function of $k$. Here $V_{0}=1$, $s_{1}=0.5$ and $y=9.5$. Transmission probability nearly vanishes for $\vert \xi_{2}\vert >0$.Horizontal dotted lines correspond to the vertical axis at $-1$ and $1$.}
\label{TN1N2_xi2_k}
\end{center}
\end{figure}

We expand $\xi_{2}$ near $k\sim k^{*}$ upto first order in $(k-k^{*})$ and using $U_{N_{1}-1}(\xi_{}^{*})=0$ we have the expression of $\xi_{2}$ near $k^{*}$ as
\begin{equation}
\xi_{2}(k)=-U_{N_{1}-2} (\xi_{1}^{*})\cos{k^{*}y}+ \big[(\tilde{F_{1}^{*}}+ \tilde{F_{2}^{*}}\sqrt{1+(\beta^{*})^{2}} \cos\zeta^{*}) Z^{*}+yU_{N_{1}-2} (\xi_{1}^{*})\sin{k^{*}y} ] (k-k^{*})
\label{xi2_taylor}
\end{equation}
Where
\begin{equation}
\tilde{F_{1}}=\frac{1}{3}(N_{1}-2)(N_{1}-1)N_{1}\cos{ky} F_{2} ^{1} (-N_{1}+3,N_{1}+1,\frac{5}{2};\frac{1}{2} (1-\xi_{1}^{*}))
\label{F1_tilde}
\end{equation}  
\begin{equation}
\tilde{F_{2}}=\frac{1}{3}(N_{1}-1)N_{1}(N_{1}+1) F_{2} ^{1} (-N_{1}+2,N_{1}+2,\frac{5}{2};\frac{1}{2} (1-\xi_{1}^{*}))
\label{F2_tilde}
\end{equation}
\begin{equation}
Z=-\beta s_{1}\cos{ks_{1}}+\frac{\sin{k s_{1}}}{k} (\beta +k s_{1})
\label{ZZ}
\end{equation}   
\begin{equation}
\zeta=\tan^{-1}\beta+k((2N_{1}-1)s_{1}+y).
\label{zetaa}
\end{equation}
In Eq. \ref{xi2_taylor} the superscript $*$ denotes the expressions evaluated at $k=k^{*}$. From \ref{xi2_taylor} the approximate value of resonance width is derived as
\begin{equation}
\Delta k=\frac{2+U_{N_{1}-2} (\xi_{1}^{*})\cos{k^{*}y}}{(\tilde{F_{1}^{*}}+\tilde{F_{2}^{*}} \sqrt{1+(\beta^{*})^{2}} \cos\zeta^{*}) Z^{*}+yU_{N_{1}-2} (\xi_{1}^{*})\sin{k^{*}y} }.
\label{resonance_width}
\end{equation}
$\Delta k$ value evaluated from Eq. \ref{resonance_width} is $0.011692$ for Fig. \ref{T22_T240_resonance_peaks} while the numerical value is found to be $0.011928$.
\paragraph{Effect of periodic distance $c_{2}$:}
From Eq. \ref{xi2_periodic_dirac_comb_other_expression} we see that the periodic distance $c_{2}$ affect the transmission probability through $y=c_{2}-s_{1}$. The quantity $ky$ only appears as argument of trigonometric functions in expression of $\xi_{2}$ and therefore with varying $y$ , $\xi_{2}$ oscillates between positive and negative real numbers. For a given $y$ , $\xi_{2}$ admits a number of zeros between the energy ranges over which perfect transmission occurs for the Dirac comb.  The zeros of $\xi_{2}$ occur if following transcendental equation is satisfied,
\begin{equation}
\sqrt{(1+\beta^{2})} U_{N_{1}-1} (\xi_{1}) \cos{\zeta (k)}=U_{N_{1}-2} (\xi_{1}) \cos{ky}
\label{xi2_zero_condition}
\end{equation}
For even $N_{2}$, every $k$ at which $\xi_{2}$ vanishes is the energy of transmission resonance as $U_{N_{2}-1} (0)$. It must be noted that due to oscillatory nature of $\xi_{2}$, a portion of $\xi_{2}-k$ curve lies in the range $ [-1,1]$ in the neighbourhood of $\xi_{2}(k)=0$. This implies $N_{2}-1$ simple roots of $U_{N_{2}-1}$ near zeros of $\xi_{2}$. Therefore the `periodic Dirac comb' will have a number of resonance bands each containing $N_{2}-1$ resonance peaks between the region of energy over which the transmission resonance occurs for the corresponding `Dirac comb'. The number of such resonance band depends upon value of $y$. We illustrate the emergence of resonance band in Fig. \ref{TN1N2_with_y}. For $\vert \xi_{2} \vert >1$, and for large $N_{2}$, $ U_{N_{2}-1} (\xi_{2}) \rightarrow \infty$. This corresponds to the forbidden energies of the periodic Dirac comb and subsequently the band structure. Vanishingly small transmission probability for $\vert \xi_{2} \vert >1$ is demonstrated in the Fig. \ref{TN1N2_xi2_k}.
\subsection{Super periodic rectangular potential}
If the unit cell is a rectangular barrier of height $V$ confined over a distance $0 \leq  x \leq b$ then we have a super periodic rectangular potential. The transfer matrix of the rectangular barrier is \cite{griffith_pp}
\begin{equation}
 M= \begin{pmatrix}   (\cos{k'b}-i\varepsilon_{+} \sin{k'b})e^{ikb} & i\varepsilon_{-} \sin{k'b} \\ -i\varepsilon_{-} \sin{k'b} & (\cos{k'b}+i\varepsilon_{+} \sin{k'b})e^{-ikb}   \end{pmatrix}  
\label{transfer_matrix_rectangular_potential}
\end{equation} 
where
\begin{equation}
\varepsilon_{\pm}=\frac{1}{2} \Big( \mu\pm \frac{1}{\mu}\Big)  
\label{epsilon_plus_minus}
\end{equation}
\begin{equation}
\mu=\frac{k}{k'}  \nonumber
\end{equation}
and $k=\sqrt{2mE}/\hbar $ , $k'=\sqrt{2m(E-V)}/\hbar$.
Therefore for rectangular potential the modulus of $m_{22}$ term of transfer matrix \ref{transfer_matrix_rectangular_potential} is
\beq
\vert m_{22}\vert=(\cos^{2}{k'b}+\varepsilon_{+}^{2}\sin^{2}{k'b})^{\frac{1}{2}} 
\label{abs_m22_rectangular_potential}
\eeq
and the argument is
\beq
\alpha=\tan^{-1}{(\varepsilon_{+} \tan{k'b})}-kb
\label{alpha_rectangular_barrier}
\eeq
This is all we need to calculate various $\xi$'s from Eq. \ref{xin}. The transmission probability for the super periodic rectangular potential of any order $n$ can be easily found from Eq. \ref{transmission_2} with $\vert m_{12}\vert =\varepsilon_{-}\sin{kb}$. Scattering from the periodic rectangular barrier has been discussed in details (see references in \cite{griffith_pp}).  For the purpose of completeness, below we give the expression of transmission probability $T(N_{1};k)$ for the periodic rectangular barrier. 
\beq
T(N_{1};k)=\frac{1}{1+[\varepsilon_{-}\sin{(kb)} U_{N_{1}-1}(\xi_{1})]^{2}} 
\label{t_periodic_rectangular}
\eeq
where,
\beq
\xi_{1}=(\cos^{2}{k'b}+\varepsilon_{+}^{2}\sin^{2}{k'b})^{\frac{1}{2}} \cos{[\tan^{-1}{(\varepsilon_{+} \tan{k'b})}+kc_{1}]}
\eeq
Now we discuss the special case which is  super periodic rectangular potential of order $2$.
\subsubsection{Super periodic rectangular potential of order $2$}
\begin{figure}
\begin{center}
\includegraphics[scale=0.6]{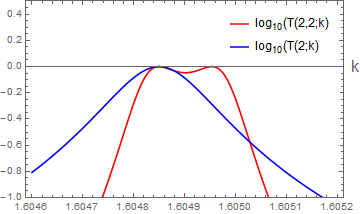} 
\caption{Transmission probability $T(2;k)$ and $T(2,2;k)$ as a function of $k$ for $V=10$, $b=1.0$, $c_{1}=1.5$, $c_{2}=2.0$ near the transmission resonance of $T(2,k)$. The resonance point is $k^{*}=1.60485$. Note the appearance of $N_{2}=2$ resonance peaks near $k^{*}$  }
\label{resonance_peak_sprp}
\end{center}
\end{figure}

\begin{figure}
\begin{center}
\includegraphics[scale=0.5]{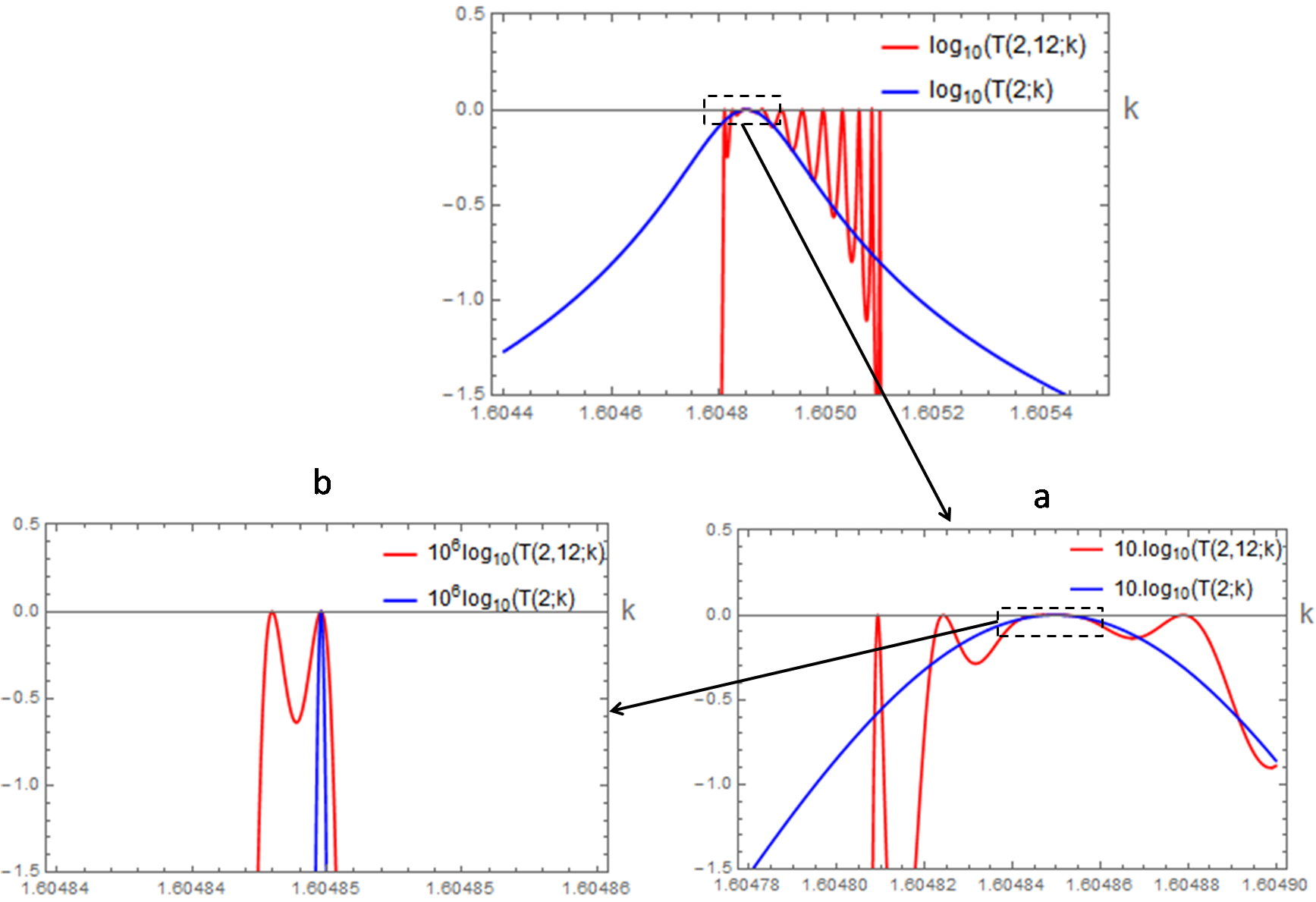} 
\caption{Transmission probability $T(2;k)$ and $T(2,12;k)$ as a function of $k$ for the same parameters as Figure \ref{resonance_peak_sprp}. There are total $N_{2}=12$ number of resonance peaks near $k=k^{*}$. The resonance peaks in the immediate neighbourhood of $k^{*}$ is difficult to resolve from the figure. In Fig. -$a$ above, $\log_{10} T$ is amplified $10$ times while in Fig. -$b$ above, it is $10^{6}$ times amplified to show the resonance features.}
\label{resonance_peak_sprp_12}
\end{center}
\end{figure}

The transmission probability for super periodic rectangular potential of order $2$ can be easily written as
\beq
T(N_{1},N_{2};k)=\frac{1}{1+[(\varepsilon_{-}\sin{k'b}) U_{N_{1}-1}(\xi_{1})U_{N_{2}-1}(\xi_{2})]^{2}}
\label{TN1N2_rectangular_barrier}
\eeq
where,
\beq
\xi_{2}=(\cos^{2}{k'b}+\varepsilon_{+}^{2}\sin^{2}{k'b})^{\frac{1}{2}} U_{N_{1}-1} (\xi_{1}) \cos{(\tan^{-1}{(\varepsilon_{+}\tan{k'b})}-kc_{2})}-U_{N_{1}-2} (\xi_{1})\cos{k(c_{1}-c_{2})}
\label{xi2_rectangular_barrier}
\eeq
and $c_{1}$ and $c_{2}$ are the periodic distances defined earlier.
Again it is evident that the transmission resonance of the periodic rectangular potential is also the transmission resonances of super periodic potential of order $2$. Further it is easy to work out that the presence of super periodicity of second order gives rise to a resonance band containing $N_{2}-1$ peaks in the neighbourhood of the transmission resonances of the periodic rectangular potential. The reason of this has been discussed in section \ref{periodic_dirac_comb}. The overall effect is that the transmission resonances of the periodic rectangular potential convert to a resonance band containing $(N_{2}-1)+1=N_{2}$ resonance peaks over the energy range for which $\xi_{2} \in [-1,1]$. We show this graphically in Fig. \ref{resonance_peak_sprp} and Fig. \ref{resonance_peak_sprp_12}. We also found that the width of the resonance band is the span of $k$ over which $\xi_{2} \in [-1,1]$. This is the similar result as obtained in the case of the periodic Dirac comb.

\paragraph{}

Next we discuss the applications of super periodic formalism developed here so far. We apply the super periodicity of rectangular potential to derive the close form expressions of fractal potentials. We find that few of the fractal potentials of the family  of Cantor sets are special cases of the super periodic rectangular potential. This shows the super periodic set is a more general set than certain fractals set. This needs further investigation to examine the class of fractals that are the subset of super periodic operations.      
\subsection{Cantor fractal potential}
The Cantor set is one of the simplest example of fractals. In this section we show that some family of Cantor set potentials are special cases of super periodic rectangular potentials. We discuss three types of Cantor set potentials: standard Cantor set, general Cantor set and Smith-Volterra Cantor set potentials. The close form expressions of the transmission amplitude are obtained by using the concept of super periodicity.
\subsubsection{Standard Cantor potential} 
\begin{figure}
\begin{center}
\includegraphics[scale=0.5]{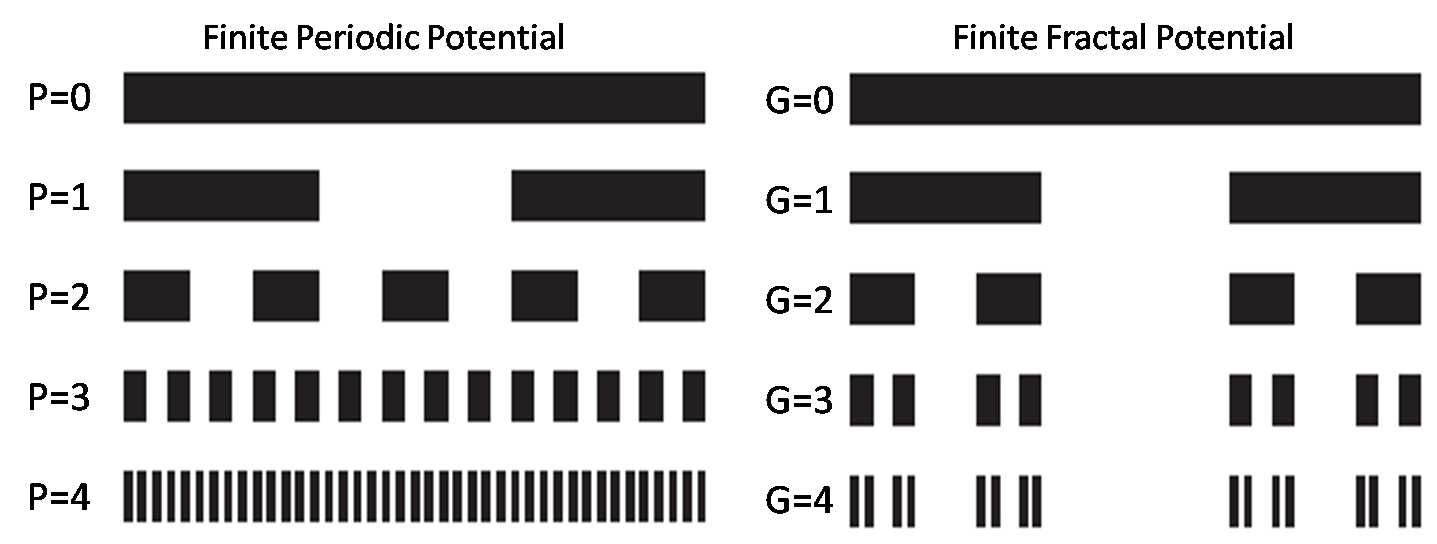} 
\caption{Finite periodic (left) and Cantor set (right) potentials. Here the white region shows the gap between the potential and the height  of black region is the potential height $V$}
\label{figure_fractal_potential}
\end{center}
\end{figure}
Standard Cantor set is the simplest fractals in the family of Cantor like fractals. In general a stage $G$ Cantor set potential (of height $V$) with the total span $L$ is obtained by dividing the length $L$ in three equal parts iteratively and removing the central portion each time as shown in Fig. \ref{figure_fractal_potential}. At stage $G=0$, the potential is a rectangular potential of height $V$ and width $L$. In Fig. \ref{figure_fractal_potential} Cantor potential of stage $G=4$ is shown. At any stage $G$, the potential consists of $2^{G}$ rectangular potentials each of width $l_{G}=\frac{L}{3^{G}}$. 
\paragraph{}
It is evident that a Cantor potential of stage $G$ is the super periodic rectangular potential of order $G$ where the unit cell rectangular potential is of height $V$ and width $b=l_{G}$ with $N_{i}=2$ and 
\beq
s_{i}=2 \big ( \frac{L}{3^{G+1}} \big)3^{i}
\label{si_Cantor}
\eeq
for $i \in \{1,2,3,..,G\}$. Therefore from Eq. \ref{xin} one can obtain $\xi_{j}$. After the algebra, the simplified expression of $\xi_{j}$ is 
\beq
\xi_{j}=2^{j-1}\vert m_{22} \vert \cos{(\alpha +\tau(3+3^{j}))} \prod_{i=1}^{j-1}\xi_{i}-   \sum_{p=1}^{j-1} \Big ( 2^{j-p-1}\cos{(3^{p}-3^{j})\tau}  \prod_{i=p+1}^{j-1} \xi_{i} \Big)  -\cos{(2\tau 3^{j-1})}
\label{xi_Cantor}
\eeq
Where,
\beq
\tau(k)=\frac{kL}{3^{G+1}}
\eeq
and the expression for $\vert m_{22}\vert$  and $\alpha$ are given by Eqs. \ref{abs_m22_rectangular_potential} and \ref{alpha_rectangular_barrier} respectively with $b=l_{G}=\frac{L}{3^{G}}$. 
In Eq. \ref{xi_Cantor} it must be noted that when $p > j-1$, the summation term will be dropped. In deriving the Eq. \ref{xi_Cantor} we have used $U_{0}(\xi)=1$ and $U_{1}(\xi)=2\xi$. The transmission probability of Cantor set potential of stage $G$ is easily obtained as
\beq
T_{G}(k)=\frac{1}{1+4^{G}\varepsilon_{-}^{2}\sin^{2}{(k'l_{G})} \prod_{i=1} ^{G} \xi_{i}^{2}}
\label{T_Cantor}
\eeq
\subsubsection{General Cantor potential}
\begin{figure}
\begin{center}
\includegraphics[scale=0.5]{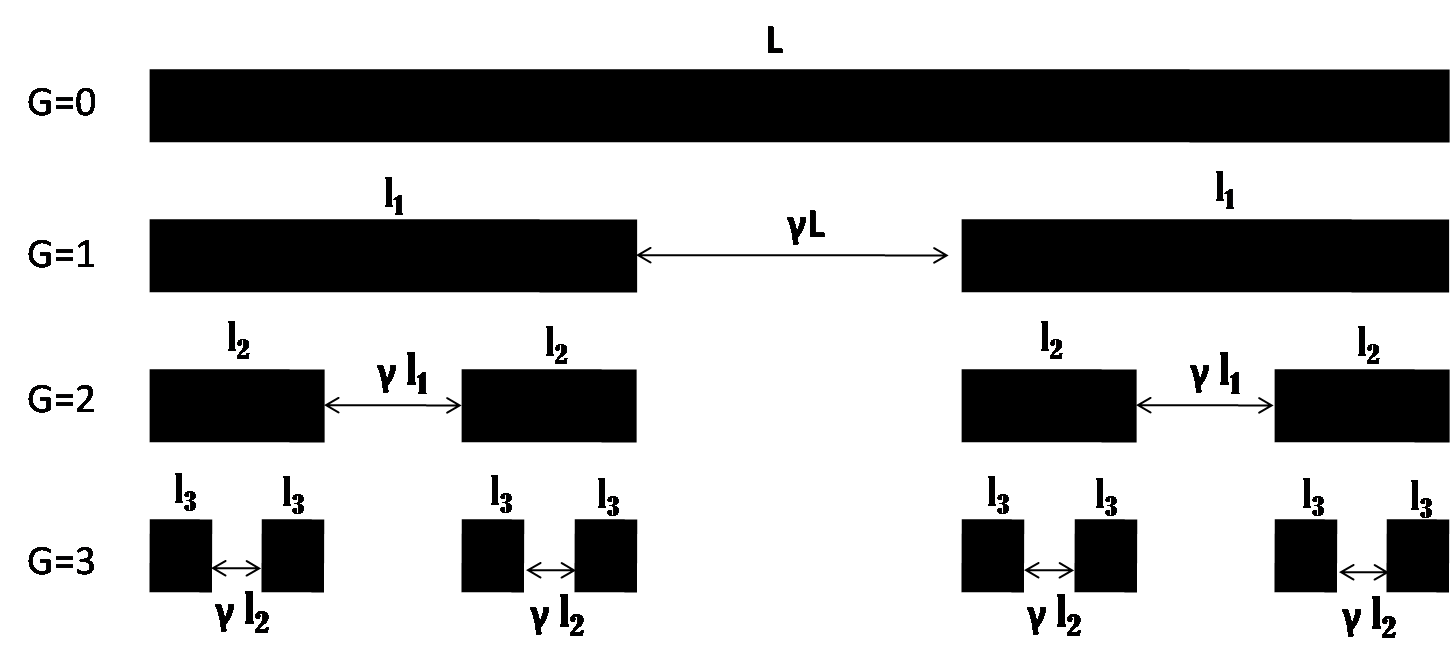} 
\caption{The general Cantor set potential.Here the white region shows the gap between the potential and the height $V$ of black region is the poetntial height}
\label{general_cantor_potential}
\end{center}
\end{figure}

\begin{figure}
\begin{center}
\includegraphics[scale=0.30]{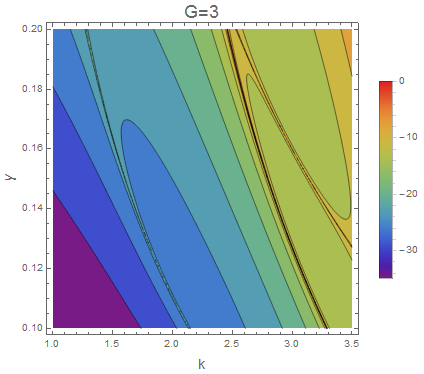} a \ \includegraphics[scale=0.30]{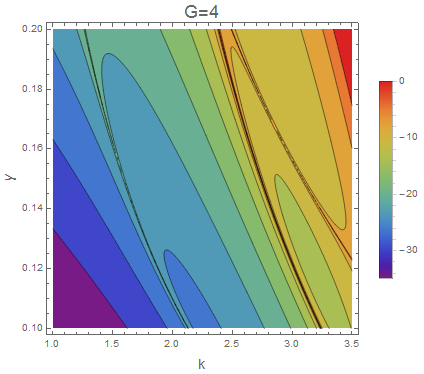} b \ \includegraphics[scale=0.30]{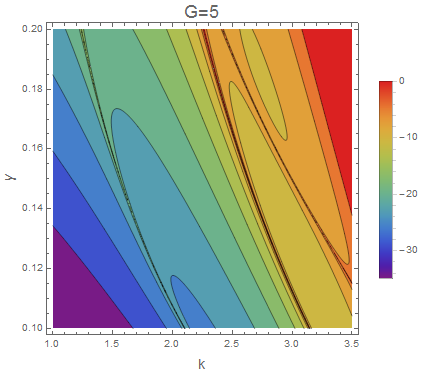} c \\
\includegraphics[scale=0.30]{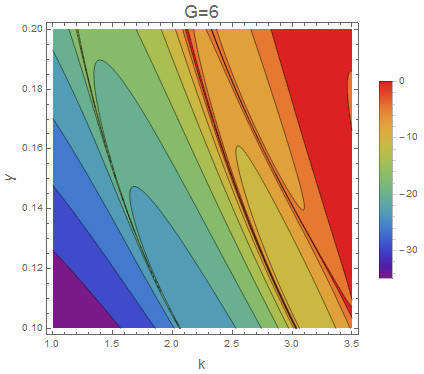} d \  \includegraphics[scale=0.30]{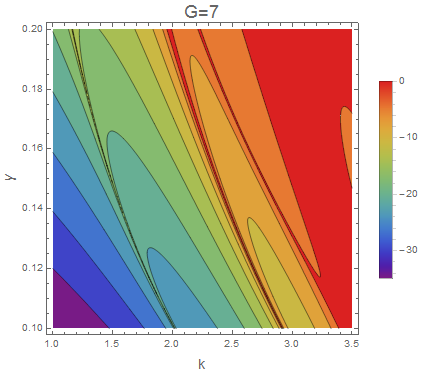} e \ \includegraphics[scale=0.30]{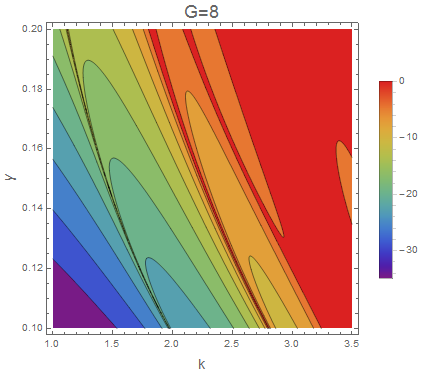} f \\
\includegraphics[scale=0.30]{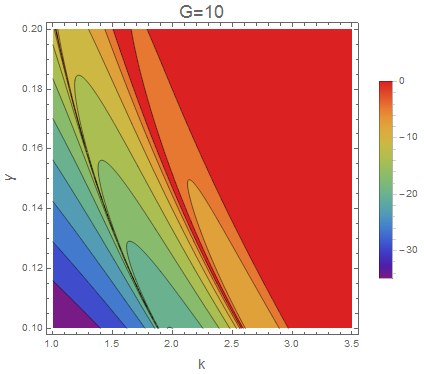} g \  \includegraphics[scale=0.30]{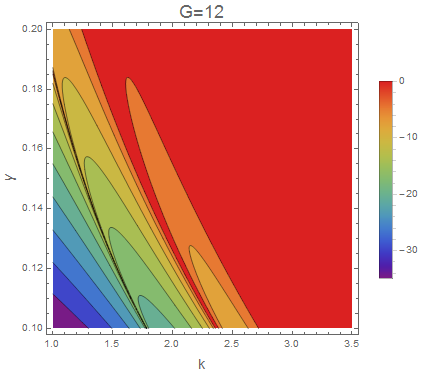} h \ \includegraphics[scale=0.30]{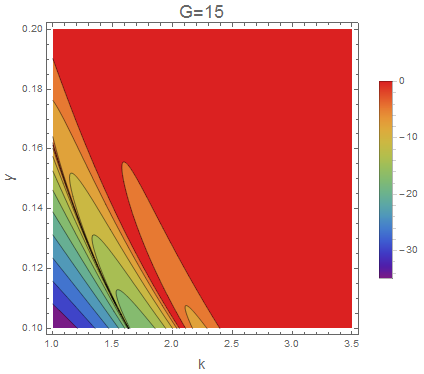} i \\
\end{center}
\caption{Contour plot of $\log_{10}T$ on $\gamma -k$ plane for general Cantor potential. The color bar for each figure is shown adjacent to the figure which varies from $-35$ (violet) to $0$ (red). Here $V_{0}=10$ and $L=10$.}
\label{cantor_contour_plot}
\end{figure}

The general Cantor set potential is shown in Fig \ref{general_cantor_potential}. In general a stage $G$ general Cantor set potential (of total span $L$) is obtained by removing $G$ times a fraction $\gamma$ of the length from the middle portion of each segment iteratively. For stage $G=3$, this is illustrated in the Fig \ref{general_cantor_potential}. Again at stage $G=0$, the potential is a rectangular potential of height $V$ and width $L$. At stage $G$, the potential consists of $2^{G}$ rectangular potential each of width 
\beq
l_{G}=\big (\frac{1-\gamma}{2}\big)^{G}L
\label{lg_general_cantor}
\eeq
 \
For $\gamma=\frac{1}{3}$ one can obtain the standard Cantor potential. Similar to Cantor  potential it is observed that a general Cantor potential of stage $G$ is also a super periodic rectangular potential of order $G$ where the width of unit cell rectangular potential is given by Eq. \ref{lg_general_cantor}. In this case $N_{i}=2$ and it can be shown that 
\beq
s_{i}=\big (\frac{1-\gamma}{2}\big)^{G-i} \big( \frac{1+\gamma}{2}\big) L
\eeq
for $i \in \{1,2,3,..,G\}$. Again the general expression for $\xi_{j}$, $j \in \{1,2,3,..,G\}$ can be calculated from Eq. \ref{xin}. This turns out to be
\begin{multline}
\xi_{j}=2^{j-1}\vert m_{22} \vert \cos{\Big\{\alpha +\tau_{\gamma} \Big( \big( \frac{1-\gamma}{2}\big)^{-1} +\big(\frac{1-\gamma}{2}\big)^{-(j+1)}\gamma \Big)\Big\} } \prod_{i=1}^{j-1}\xi_{i}-  \\
 \sum_{p=1}^{j-1}  \Big [ 2^{j-p-1}  \cos{\Big\{ \tau_{\gamma} \gamma \Big( \big( \frac{1-\gamma}{2}\big)^{-(p+1)}-\big( \frac{1-\gamma}{2}\big)^{-(j+1)} \Big)\Big\}} \prod_{i=p+1}^{j-1} \xi_{i} \Big] \\ -\cos{\Big\{\tau_{\gamma} \gamma \big(\frac{1+\gamma}{2}\big) \big(\frac{1-\gamma}{2}\big)^{-(j+1)} \Big\}}
\label{xi_general_cantor}
\end{multline}
where,
\beq
\tau_{\gamma}=kL \Big(\frac{1-\gamma}{2} \Big)^{G+1}
\label{tau_gamma}
\eeq
For $\gamma=\frac{1}{3}$, Eq. \ref{xi_general_cantor} reduces to \ref{xi_Cantor}. Now the transmission probability from general Cantor potential of stage $G$ is computed as follows:
\beq
T_{G}(k)=\frac{1}{1+4^{G}\varepsilon_{-}^{2}\sin^{2}{(k'l_{G})} \prod_{i=1} ^{G} \xi_{i}^{2}}
\label{T_general_cantor}
\eeq
The contour plot of transmission probability with varying $\gamma$ and $k$ is shown for general Cantor potential of different stage $G$ in Fig. \ref{cantor_contour_plot}. These plot shows very sharp features of the transmission resonances as well as null transmissions of the wave. As $G$ increases, the unit cell rectangular potential becomes more thinner and the wave easily transmits the fractal system. This is evident from the plots in Fig \ref{cantor_contour_plot} which shows that as $G$ increases the red region slowly sweeps from larger $k$ to lower $k$ region in $\gamma - k$ plane.     

\subsubsection{Smith-Volterra-Cantor potential}
The construction of Smith-Volterra-Cantor (SVC) potential of total span $L$ is similar to the construction of Cantor  potential of total span $L$. It begins by removing the middle $\frac{1}{4}$ from the length $L$ (stage $G$=1). In the next step further middle $\frac{1}{4}$ from the two segments is removed (stage $G$=2). Therefore at any stage $G$ a fraction $\frac{1}{2^{2G}}$ is removed from the middle of each $2^{G-1}$ segments. At any stage $G$ there are total $2^{G}$ segments of width
\beq
l_{G}=\frac{2^{G}+1}{2^{2G+1}}L
\label{lg_svc}
\eeq 
\begin{figure}
\begin{center}
\includegraphics[scale=0.5]{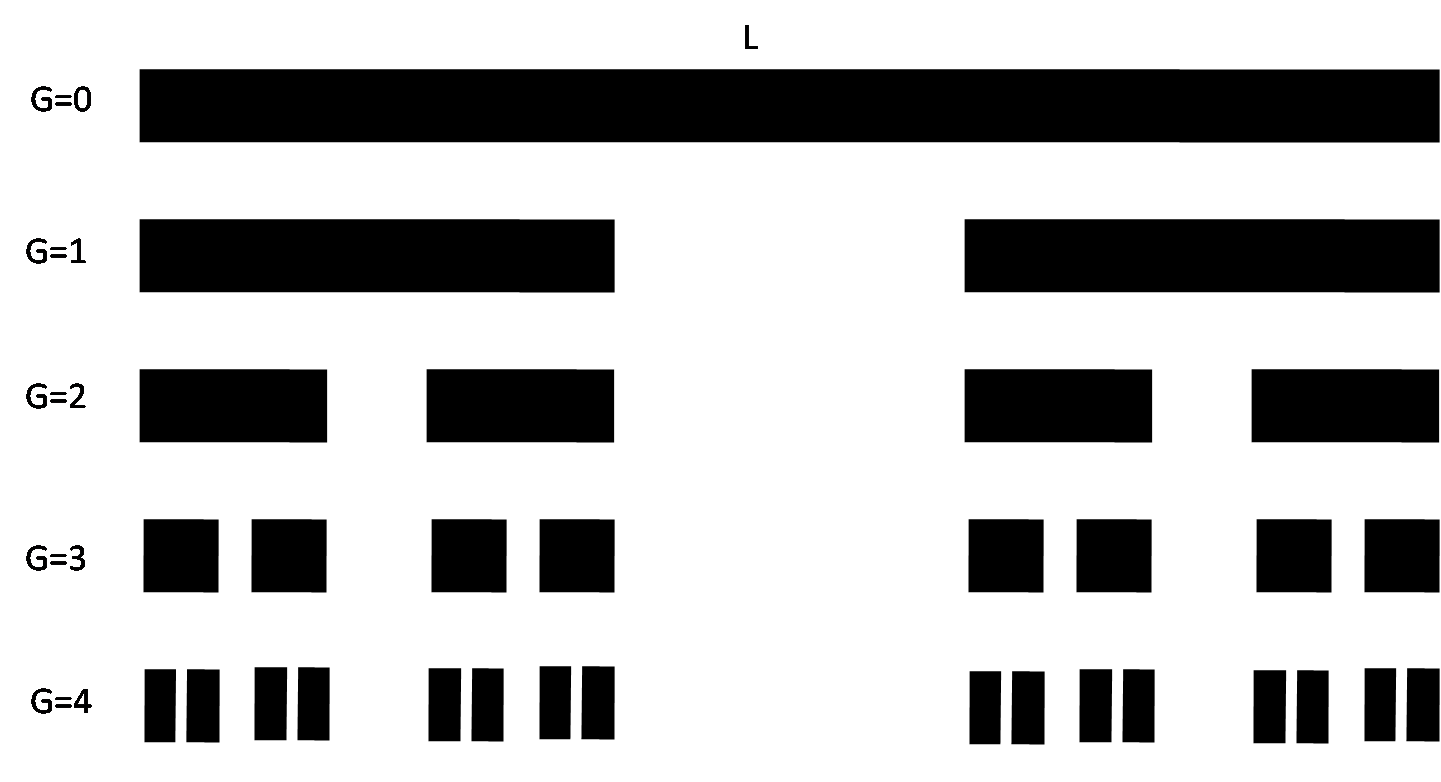} 
\caption{The Smith-Volterra Cantor potential. Here the white region shows the gap between the potential and the height of black region is the potential height  $V$. Note that consecutively lesser fraction of the previous segment is removed at a stage $G$.}
\label{svc_figure}
\end{center}
\end{figure}
The SVC potential is shown in Figure \ref{svc_figure}. Again it is easy to see that the SVC potential of stage $G$ is the super periodic rectangular potential of order $G$ with $N_{i}=2$, $i \in \{ 1,2,3,...,G\}$. The width of the unit cell rectangular potential is given by Eq. \ref{lg_svc}. The calculation of $s_{i}$, $i \in \{ 1,2,3,...,G\}$ is illustrated below.
\begin{figure}
\begin{center}
\includegraphics[scale=0.35]{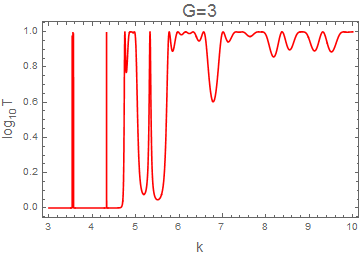} a \ \includegraphics[scale=0.35]{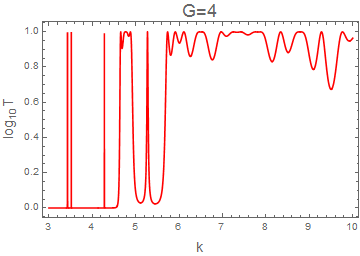} b \ \includegraphics[scale=0.35]{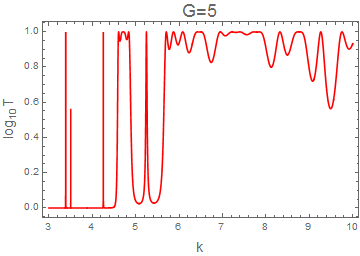} c \\
\includegraphics[scale=0.35]{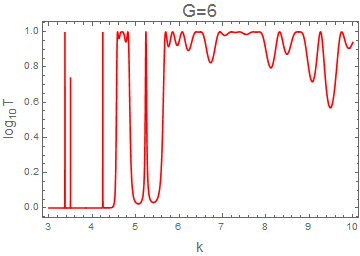} d \  \includegraphics[scale=0.35]{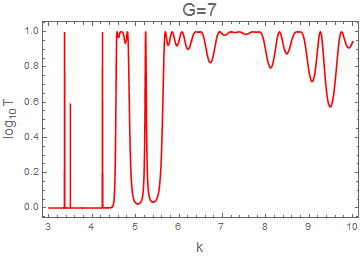} e \ \includegraphics[scale=0.35]{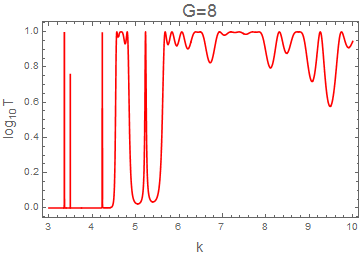} f \\
\caption{ The transmission probability for SVC potential of different stage $G$. The parameters of the potential are same as of Fig. \ref{cantor_contour_plot}. Note that with increasing $G$, the transmission profile is nearly identical.  }
\label{cantor_svc_contour_plot}
\end{center}
\end{figure}

\begin{figure}
\begin{center}
\includegraphics[scale=0.70]{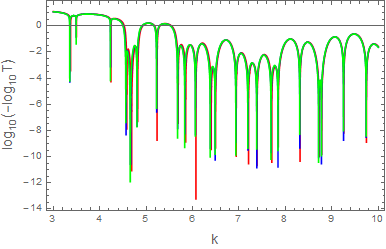}  
\label{svc_log}
\caption{Plot of $\log_{10}(-\log_{10}T)$ for the SVC potential for $G=6$ (red curve), $G=8$ (blue curve) and $G=12$ (green curve). The parameters of the potential are same as of Fig. \ref{cantor_contour_plot}. Nearly all curves overlap which indicates the tunnelling saturation with increasing $G$ }
\end{center}
\end{figure}

\paragraph{}
For the SVC potential at stage $G$, a fraction of $\frac{1}{2^{2G}}$ is removed from the middle of segment of length $l_{G-1}$. Hence 
\beq
s_{1} = l_{G}+\frac{l_{G-1}}{2^{2G}}  \nonumber
\eeq
\beq
s_{2} = l_{G-1}+\frac{l_{G-2}}{2^{2(G-1)}}   \nonumber
\eeq
\beq
s_{3} = l_{G-2}+\frac{l_{G-3}}{2^{2(G-2)}}   \nonumber
\eeq
Therefore for a general $s_{i}$ we have the following expression,
\beq
s_{i} =  l_{G+1-i}+\frac{l_{G-i}}{2^{2(G+1- i)}} \nonumber
\eeq
Using Eq. \ref{lg_svc} the expression of $s_{i}$ is simplified to
\beq
s_{i}=\frac{L}{2^{(3+4G)}}2^{i} (2^{3G+1} +2^{2G+i}+2^{G+2i}+2^{3i})
\label{si_svc}
\eeq
On knowing $s_{i}$ and $N_{i}=2$, $i \in \{ 1,2,3,...,G\}$, the general expression for $\xi_{j}$, $j \in \{ 1,2,3,...,G\}$ is obtained as
\beq
\xi_{j}=2^{j-1}\vert m_{22} \cos{[\alpha+ k \eta_{1}(j)]} \prod_{i=1}^{j-1}\xi_{i} - \sum_{p=1}^{j-1} \Big( 2^{j-p-1}  \cos{k\eta_{2}(j,p)} \prod_{i=p+1}^{j-1} \xi_{i} \Big)-\cos{k \eta_{3}(j)}
\label{xi_svc}
\eeq
Where,
\beq
\eta_{1}(j)=\frac{L}{2^{4G+3}}\frac{1}{105} \big [ 224+15. 2^{G+4} +35. 2^{2G} (2^{3}+2^{2j})-210. 2^{3G} (2^{j}-2^{2}) +2^{3j} (75.2^{G}+91.2^{j})\big]
\label{eta1}
\eeq
\beq
\eta_{2}(j,p)=\frac{L}{2^{4G+3}}\frac{2}{105} \big[ 45. 2^{G} (2^{3j}-2^{3p}) -35. 2^{2G} (2^{2j}-2^{2p})-49 (2^{4j}-2^{4p})\big ]
\label{eta2} 
\eeq
and, 
\beq
\eta_{3}(j)=\frac{L}{2^{4G+3}} 2^{2j-3} (3.2^{G+1+j}+2^{2(G+1)}+7.2^{2j})
\label{eta3}
\eeq
The transmission probability of SVC potential of any stage $G$ can be computed from the following expression,
\beq
T_{G}(k)=\frac{1}{1+4^{G}\varepsilon_{-}^{2}\sin^{2}{(k'l_{G})} \prod_{i=1} ^{G} \xi_{i}^{2}}
\label{T_svc}
\eeq
where $l_{G}$ is given by Eq. \ref{lg_svc} and various $\xi$'s are obtained from Eq. \ref{xi_svc}.
The plots of transmission probability for the SVC potential of different stages is shown in Fig \ref{cantor_svc_contour_plot}.    As consecutively lesser fraction of the previous segment is removed at a stage $G$, the transmission probability would show a saturation with increasing stage of SVC fractal potential. This is evident from the plots in Fig \ref{cantor_svc_contour_plot}. For $G=6,7 \ \mbox{and}\  8$, the profile of transmission probability are nearly the same in appearance. This is further illustrated in the Fig. \ref{svc_log}. The plots of transmission probability for $G=6, 8 \ \mbox{and} \ 12$ nearly overlap, indicating the saturation of tunnelling profile with increasing $G$ for SVC potential. 
\paragraph{}
Many other properties of tunnelling from fractal potentials such as self similarity of tunnelling probability have already been reported in the literature \cite{f1}-\cite{f8} and hence we are not discussing these things here.

\section{Results and Discussions}
\label{results_discusssions}
In this paper we have introduced the most general form of periodicity. We have studied the tunnelling behaviour of a Schrodinger particle by constructing the super periodic structure of potential. The super periodic structure has internal periodicity at every levels. The close form expressions of the elements of the transfer matrix as well as the transmission probability have been obtained in their nice closed form for SPP of arbitrary order $n$.  

We have shown analytically that if perfect transmission exists for any potential at energy $E^{*}$, then the super periodic structure of any order $n$ of that potential will also have perfect transmission at energy $E^{*}$. This also applies to each subsequent periodic entity of the super periodic set. The emergence  of band structure of super periodic potentials are modulated by super periodicity by the appearance of resonance bands. We have also discussed in detail the tunnelling behaviour for special cases of  second order SPP  with delta and rectangular potentials as `unit cell potential'.  For the  periodic Dirac comb, the band structure emerges even for $T(2,3;k)$. We have also demonstrated that the transmission resonance of periodic rectangular potential splits into a resonance band due to super periodicity. The analytical reason of the emergence of resonance bands has been discussed in details in terms of the properties of the Chebyshev polynomials. By Taylor expanding the phase  $\xi_{2}$ near resonance point to first order, an approximate formula for calculating the width of the resonance band is also given for $T(N_{1},N_{2};k)$.  

As an application of super periodic concept we  have identified a family of symmetric Cantor potential , the general Cantor potential and the Smith-Volterra-Cantor potential of stage $G$ as the special case of super periodic rectangular potential of order $G$. This is an extraordinary fact to recognise a family of symmetric fractal potential as special case of super periodicity. This connection shows that the super periodic formalism developed here is more general and not  only limited to periodicity but also to the self-similarity. This further broaden the applications and scope of the SPP formalism developed in this work.

The close form expressions of tunnelling probability (amplitude) are obtained easily using our results for those fractal potentials which are the special cases of SPP. We have calculated the close form expressions of tunnelling probability for general Cantor potential and SVC potential. It is found that with increasing stage of SVC potential the tunnelling profile $T(k)$ with $k$ saturates and the whole profile approaches a limiting value. However this limiting value will depend upon the potential height and the total span of the SVC potential $L$. For general Cantor potential, each rectangular potential segment gets thinner with increasing $G$ and thus the wave has more transmission with increasing $G$. 

We hope that the SPP formalism developed here will help to analytically understand the transmission characteristics through more sophisticated fractals and periodic super-lattices and the design of advanced micro-structures.  This would be further interesting to study the tunneling time problem analytically for SPP and general fractal potentials which have not been discussed so far. Based on the SPP formalism, the study of band structures for periodic fractal potential will be attempted in our future work.     \\
\\
{\it \bf{Acknowledgements}}: \\
MH acknowledges support from SAG/ISAC and encouragement from Dr. M. Annadurai, Director, ISAC to carry out this research work. BPM acknowledges the support from CAS, Department of Physics, BHU.

\end{document}